\documentclass[12pt,a4paper,oneside]{article}
\pdfoutput=1
\usepackage{jheppub}

\usepackage{verbatim}
\usepackage{mathrsfs}
\usepackage{appendix}
\usepackage{caption}
\usepackage{float}
\usepackage{subfig}
\usepackage[vcentermath]{youngtab}
\usepackage{alltt}
\usepackage{multirow}
\usepackage{arydshln}
\usepackage{lscape}
\usepackage{tabu}
\usepackage{amssymb,amsmath,amsfonts,mathrsfs,enumerate,wrapfig,latexsym,wasysym}
\usepackage[displaymath, mathlines]{lineno}
\usepackage{graphicx}
\usepackage{adjustbox}
\usepackage{xcolor}
\usepackage{slashed}    
\usepackage{url}
\usepackage{hyperref}
\hypersetup{
	colorlinks = true,
	linkcolor = blue,
	anchorcolor = blue,
	citecolor = blue,
	filecolor = blue,
	urlcolor = magenta,
	bookmarks=true,
	linktocpage=true
}
\usepackage{framed}
\usepackage[numbers,sort&compress]{natbib}
\usepackage{fancyvrb}
\allowdisplaybreaks
\usepackage{multirow}
\usepackage{framed}
\RequirePackage{flushend}
\usepackage{makecell}
\usepackage{color}
\usepackage{pifont}
\usepackage{cancel}
\usepackage{import}
\usepackage{subfiles}
\usepackage{filecontents}
\usepackage{rotating}
\usepackage{threeparttable,tablefootnote}\usepackage{mathtools}
\usepackage{pbox}
\usepackage{setspace}
\usepackage{lscape}
\usepackage{color, colortbl}
\usepackage{pdflscape}
\usepackage{afterpage}
\usepackage{fontawesome}

\usepackage{threeparttable}
\usepackage{slashed}
\usepackage[normalem]{ulem}

\definecolor{darkgreen}{cmyk}{1,0,1,0.4}

\definecolor{pink}{cmyk}{0.4,1,0.3,0}

\def\com2#1{\textcolor{red}{\it{#1}}}

\graphicspath{{plots/}}

\definecolor{Gray}{gray}{0.9}

\title{Primordial Monopoles and Strings, Inflation, and Gravity Waves}

\author[1]{Joydeep Chakrabortty,}

\author[2]{George Lazarides,}

\author[1]{Rinku Maji,}

\author[3]{Qaisar Shafi}

\affiliation[1]{Indian Institute of Technology Kanpur, Kalyanpur, Kanpur 208016, Uttar Pradesh, INDIA}

\affiliation[2]{School of Electrical and Computer Engineering, Faculty of Engineering,
	Aristotle University of Thessaloniki, Thessaloniki 54124, Greece}

\affiliation[3]{Bartol Research Institute, Department of Physics and Astronomy,
	University of Delaware, Newark, DE 19716, USA}

\abstract{We consider magnetic monopoles and strings that appear in non-supersymmetric $SO(10)$ and 
$E_6$ grand unified models paying attention to gauge coupling unification and proton decay in a variety of symmetry breaking schemes. The dimensionless string tension parameter $G\mu$ spans the range 
$10^{-6}-10^{-30}$, where $G$ is Newton's constant and $\mu$ is the string tension. We show how intermediate scale monopoles with mass $\sim 10^{13}-10^{14}$ GeV and flux 
$\lesssim 2.8\times 10^{-16}$ ${\mathrm{cm}^{-2}\mathrm{s}^{-1}\mathrm{sr}^{-1}}$, and cosmic strings with $G\mu \sim 10^{-11}-10^{-10}$ survive inflation and are present in the universe at an observable level. We estimate the gravity wave spectrum emitted from cosmic strings taking into account inflation driven by a Coleman-Weinberg  potential. The tensor-to-scalar ratio $r$ lies between $0.06$ and $0.003$ depending on the details of the inflationary scenario.}

\begin{document}
	
	\maketitle

	\section{Introduction}
	Grand Unified Theories (GUTs) such as $SU(5)$, $SO(10)$ (more precisely $Spin(10)$), and $E_6$ predict the existence of a topologically stable \cite{tHooft:1974kcl,Polyakov:1974ek} superheavy magnetic monopole of mass  $\sim M_X/\alpha_X$, where $\alpha_X$ denotes the gauge fine structure constant at the unified scale $M_X \sim 10^{16}$ GeV. This monopole carries a single quantum of Dirac magnetic charge as well as color magnetic charge, which is related to the fact that the unbroken subgroup is $SU(3)_C \otimes U(1)_{em}/Z_3$ \cite{Daniel:1979yz,Lazarides:1982jq}. In non-supersymmetric $SO(10)$ and $E_6$ models, the symmetry breaking to the Standard Model (SM) gauge group proceeds via one or more intermediate steps which has important consequences for monopole masses and charges. For instance, the breaking of $SO(10)$ via $SU(2)_L\otimes SU(2)_R\otimes SU(4)_C$ \cite{Pati:1974yy} yields intermediate mass monopoles that carry two quanta of Dirac magnetic charge \cite{Lazarides:1980cc,Lazarides:2019xai}.  This is an important difference from $SU(5)$ because the intermediate mass monopole in $SO(10)$ with two units of magnetic charge is a few orders of magnitude lighter than the $SO(10)$ monopole carrying one unit of charge which is again superheavy. Clearly, this cannot happen in 
$SU(5)$. Similarly, in $E_6$ if the breaking occurs via $SU(3)^3$ we find intermediate mass monopoles carrying three quanta of Dirac charge \cite{Lazarides:2019xai,Shafi:1984wk,Lazarides:1986rt,Lazarides:1988wz,Kephart:2006zd,Kephart:2017esj}. The discovery of primordial monopoles with intermediate mass scales would have profound consequences for particle physics and cosmology.

The discovery of topologically stable intermediate scale cosmic strings would also have critical ramifications for the physics of the early universe and particle physics extensions of the SM. The first and most well known example of topologically stable cosmic strings appearing in GUTs is provided by $SO(10)$ \cite{Kibble:1982ae}. If the breaking of $SO(10)$ to the SM gauge group is carried out using scalar vacuum expectation values (VEVs) in the tensor representations, a $Z_2$ symmetry remains unbroken which implies the presence of topologically stable cosmic strings. Note that a direct breaking of $SO(10)$ to the SM gauge group would yield GUT scale cosmic strings which is excluded by the WMAP and Planck satellite data \cite{Dvorkin:2011aj,Ade:2013xla} as well as the limits from pulsar timing arrays (PTA) \cite{Lentati:2015qwp,Shannon:2015ect,Blanco-Pillado:2017rnf,Arzoumanian:2018saf,Buchmuller:2019gfy}. (For recent developments see Refs.~\cite{Arzoumanian:2020vkk,Ellis:2020ena,Buchmuller:2020lbh,Pol:2020igl}). In other words, we expect the breaking to proceed via intermediate steps which is favored for non-supersymmetric $SO(10)$ for other phenomenological reasons. Thus, we are interested in exploring intermediate scale cosmic strings that appear in $SO(10)$ and $E_6$ models.

In this paper, we consider two symmetry breaking chains for each of the 
non-supersymmetric $SO(10)$ and $E_6$ GUTs with two intermediate 
steps. Superheavy magnetic monopoles with one unit of Dirac magnetic charge 
are predicted in all cases along with intermediate scale monopoles with 
two units or three units of Dirac magnetic charge in $SO(10)$ or $E_6$, 
respectively. Intermediate scale cosmic strings appear in one
of the $SO(10)$ and one of the $E_6$ models. The GUT and intermediate 
scales are determined so that all the low energy data and the constraint 
from proton decay are satisfied. We merge these models with inflation
driven by a Coleman-Weinberg potential of a scalar gauge singlet 
\cite{Shafi:1983bd, Shafi:2006cs} and further restrict the model parameters by 
requiring that the data for all the 
inflationary observables are reproduced. Studying carefully the phase
transitions during which the GUT and intermediate symmetry breakings
take place, we discuss the generation and subsequent evolution of 
magnetic monopoles and cosmic strings as well as the emission of 
gravity waves from the decaying strings.    
  		
The paper is organized as follows. In Sec.~\ref{sec:RGEs}, we summarize the important features of the renormalization group equations (RGEs) for the gauge coupling constants. This section includes a brief discussion on beta functions, Abelian mixing, and the matching conditions along with the threshold corrections. In Sec.~\ref{sec:patterns}, we describe the details of the symmetry breaking chains for the GUT models and the emergence of topological defects at different stages of symmetry breaking. We also present in this section and in the \hyperref[Appendix]{Appendix} the beta coefficients associated with each of these breaking scenarios. In Sec.~\ref{sec:unification}, we perform a goodness of fit test to estimate the solutions of RGEs for each case in terms of the unification and intermediate scales which are compatible with the low energy data and the proton lifetime constraint. We discuss in 
Sec.~\ref{sec:inflation} the inflationary dynamics with a Coleman-Weinberg potential where the inflaton is a scalar GUT singlet \cite{Shafi:1983bd}, and determine the values of the model parameters that yield successful inflation. In Sec.~\ref{sec:phasetrans}, we analyze the phase transitions during inflation which are 
associated with the unification and the intermediate scales, and in Sec.~\ref{sec:monopoles}, the monopole production during the first intermediate 
phase transition is discussed. The generation of cosmic strings during the second intermediate phase transition and their subsequent evolution is presented in Sec.~\ref{sec:strings} together with the emission of gravity waves. In Sec.~\ref{sec:afterinf}, we extend our analysis to the case that the second intermediate phase transition takes place after the end of inflation, i.e. either during inflaton oscillations or after reheating. In Sec.~\ref{sec:conclusion}, we summarize our conclusions. 

\section{Renormalization Group Equations (RGEs) for Gauge Couplings}
\label{sec:RGEs}
The RGEs for the gauge couplings $g_i$ ($i=1,2,...,n$) corresponding to a generic product gauge group of the form $\mathcal{G} \equiv\mathcal{G}_1\otimes \mathcal{G}_2\otimes...\otimes \mathcal{G}_n$ can be written as (up to two loop) \cite{PhysRevLett.30.1343,Caswell:1974gg,Jones:1974mm,Jones:1981we,Machacek:1983tz,Machacek:1983fi,Machacek:1984zw}:
\begin{equation}
\mu \frac{dg_i}{d\mu} = \frac{1}{16\pi^2} b_i g_i^3 + \frac{1}{(16\pi^2)^2}\sum_{j=1}^n b_{ij} g_i^3g_j^2,
\end{equation}
where $\mu$ is the renormalization scale parameter (not to be confused with the string tension) and
\begin{eqnarray}\label{beta_coef}
b_i &= &\frac{4\kappa}{3}T(F_i)D_{F_i}
+\frac{1}{3}\eta T(S_i)D_{S_i} - \frac{11}{3} C_2(\mathcal{G}_i),  \nonumber \\
b_{ij} & = &\left[\left(\frac{20}{3} C_2(\mathcal{G}_i)+4C_2(F_i)\right)\kappa T(F_i)D_{F_i}\right. \nonumber  \\
& &\left.+ \left(\frac{2}{3} C_2(\mathcal{G}_i)+4C_2(S_i)\right)\eta T(S_i)D_{S_i} - \frac{34}{3} (C_2(\mathcal{G}_i))^2\right]\delta_{ij} \nonumber \\
& & + 4 \left(\kappa C_2(F_j) T(F_i) D_{F_i} + \eta C_2(S_j) T(S_i) D_{S_i} \right)
\end{eqnarray}
are the one- and two-loop $\beta$-coefficients respectively with $\kappa=1\;(1/2)$ for Dirac (Weyl) fermions and $\eta=1\;(1/2)$ for complex (real) scalars. $F_i$ ($S_i$) denote the fermion (scalar) representations transforming under $\mathcal{G}_i$, $T(R_i)$ is the normalization of the representation $R_i$\footnote{It is defined as $\mathrm{Tr} \left( T^a T^b \right) = T(R)\delta^{ab}=2\ell_R \delta^{ab}$, with $T^a$ being the generators of the group, $\ell_R$ is the Dynkin index  corresponding to the representation $R$, and $a,b = 1,2,\cdots, d_{\mathcal{G}}$, where $d_{\mathcal{G}}$ is the dimension of the group.}, $C_2(\mathcal{G}_i)$ is the quadratic Casimir operator for the group $\mathcal{G}_i$, and 
$C_2(R_i)$ is the quadratic Casimir operator for the representation $R_i$. Also, $D_{R_i}  =\prod_{j\neq i}D(R_j)$ with $D(R_i)$ being the dimension of the $i$th representation in the multiplet $R=(R_1,R_2, ... , R_n)$. 

The multiple occurrence of Abelian groups leads to the mixing of their gauge couplings even at the 
one-loop level \cite{Holdom:1985ag,delAguila:1988jz,Lavoura:1993ut,delAguila:1995rb,Bertolini:2009qj,Chakrabortty:2009xm,Fonseca:2013bua,Chakrabortty:2017mgi}.
In this case, instead of treating the individual evolution of each Abelian gauge coupling, we need to consider the complete Abelian gauge coupling matrix, e.g. for two Abelian gauge groups $U(1)_1\otimes U(1)_2$ we should consider the following  matrix  
\begin{equation}\label{eq:mixing-matrix}
g = \begin{pmatrix}
g_{11} & g_{12} \\
g_{21} & g_{22}
\end{pmatrix},
\end{equation}
and the RGEs of the individual matrix elements $g_{cb}$ with $c,b = 1,2$ can be expressed as:
\begin{equation}
\mu \frac{dg_{cb}}{d\mu}=\beta_{ab} g_{ca},
\end{equation}
where
\begin{equation}
\beta_{ab} = \frac{1}{(4\pi)^2} g_{ia} \left( \beta_{ij}^{\rm{1L}} + \frac{1}{(4\pi)^2}\beta_{ij}^{\rm{2L}}\right) g_{jb}.
\end{equation}
The one-loop beta coefficients are 
\begin{equation}
\beta_{ij}^{\rm{1L}} = \tilde{b}_{ij} = \frac{4}{3}\kappa q_i^F q_j^F D(F) + \frac{1}{3}\eta q_i^S q_j^S D(S),
\end{equation}
where $q_i^{F(S)}$ is the Abelian $U(1)_i$ charge of the fermion (scalar) multiplet $F\; (S)$. Similarly, the two-loop beta coefficients are 
\begin{equation}
\beta_{ij}^{\rm{2L}} = \tilde{b}_{ij,kl} g_{km}g_{lm} = \tilde{b}_{ij,kl}(g_{k1}g_{l1} + g_{k2}g_{l2}),
\end{equation}
with
\begin{equation}
\tilde{b}_{ij,kl} = 4\left( \kappa q_i^F q_j^Fq_k^F q_l^F D(F) + \eta q_i^S q_j^S q_k^S q_l^S D(S) \right).
\end{equation}
It is interesting to note that at the two-loop level, the RGEs of the non-Abelian gauge couplings $g_r$ receive additional contributions due to the Abelian gauge coupling mixing. The additional contributions are of the following forms: 
\begin{equation}
\mu \frac{dg_r}{d\mu} \supset \frac{1}{(4\pi)^4} b_{ij,r}g_r^3g_{ik}g_{jk} \;\;\;\; \text{and} \;\;\;\;
\beta_{ij}^\mathrm{2loop}\supset \tilde{b}_{ij,r} g_{r}^2\,,
\end{equation}
where
\begin{eqnarray}
b_{ij,r} &=& 4 \left(\kappa  q_i^F q_j^F T(F_r) D_{F_r} + \eta q_i^S q_j^S T(S_r) D_{S_r} \right), \nonumber \\
\tilde{b}_{ij,r} &=& 4 \left(\kappa  q_i^F q_j^F C_2(F_r) + \eta q_i^S q_j^S C_2(S_r) \right).
\end{eqnarray}

If a non-Abelian parent symmetry $\mathcal{G}_P$ is spontaneously broken to another non-Abelian daughter symmetry $\mathcal{G}_D$, the appropriate matching condition at the scale $\mu$ along with the 
one-loop threshold correction $\Lambda_D(\mu)$ is given as \cite{WEINBERG198051, Hall:1980kf,Bertolini:2009qj,Bertolini:2013vta, Chakrabortty:2019fov,Bandyopadhyay:2019rja}:
\begin{equation}\label{matching}
\frac{1}{\alpha_D(\mu)}-\frac{C_2(\mathcal{G}_D)}{12\pi}=\left(\frac{1}{\alpha_P(\mu)} -\frac{C_2(\mathcal{G}_P)}{12\pi}\right) - \frac{\Lambda_D(\mu)}{12\pi},
\end{equation}
where 
\begin{equation}\label{lamda}
\Lambda_D(\mu)=-21\; {\mathrm{Tr}} (t_{DV}^2 \ln\frac{M_V}{\mu})+2\;\eta\; {\mathrm{Tr}} (t_{DS}^2 \ln\frac{M_S}{\mu}) + 8\;\kappa\; {\mathrm{Tr}} (t_{DF}^2 \ln\frac{M_F}{\mu}),
\end{equation}
and $\alpha_i=g_i^2/4\pi$.
Here, $t_{D\psi}$ denotes the generators in the superheavy representation of $\mathcal{G}_D$ with $\psi \in \{V,S,F\}$ referring to the vector, scalar, and fermion fields respectively with masses $M_{\psi}$.
The above matching condition is modified if the daughter symmetry is an Abelian $U(1)_D$ and  originates from multiple non-Abelian parent symmetries $\mathcal{G}_{P_i}$:
\begin{equation}
\frac{1}{\alpha_D(\mu)}=\sum_i{ \omega_i^2\left(\frac{1}{\alpha_{P_i}(\mu)} -\frac{C_2(\mathcal{G}_{P_i})}{12\pi}\right)} - \frac{\Lambda_D(\mu)}{12\pi},
\end{equation} 
where the $\omega_i$'s are the weight factors of the Abelian mixing with $\sum_i{ \omega_i^2}=1$.

\section{Symmetry Breaking and Topological Defects}
\label{sec:patterns}
In this paper we study two non-supersymmetric unified theories based on each of the gauge groups 
$SO(10)$ and $E_6$ with specific symmetry breaking chains. Regarding to notation, we denote a gauge group of the form $SU(m)_A\otimes SU(n)_B\otimes U(1)_C$ as $\mathcal{G}_{m_An_B1_C}$. Any representation under this product  group is expressed as  $(p,q,r)$, which implies that it transforms as 
$p$- and $q$-dimensional representation under $SU(m)_A$ and $SU(n)_B$ respectively carrying $U(1)_C$ charge $r$.  For example, $\mathcal{G}_{2_L2_R4_C}$ stands for  $SU(2)_L\otimes SU(2)_R\otimes SU(4)_C$ and the representation depicted as  $(1,3,15)$ transforms as a singlet under $SU(2)_L$, a triplet under $SU(2)_R$, and as the adjoint representation under $SU(4)_C$. We will employ, throughout the paper, the so-called extended survival hypothesis  which states that, at each stage of symmetry breaking, the only Higgs fields which do not decouple are the ones required for the subsequent symmetry breakings.  
 
\section*{$SO(10)$ Unification through Two Intermediate Steps}

\begin{figure}[htb!]\label{fig:so10-breaking}
	\begin{center}
		\includegraphics[width=16cm]{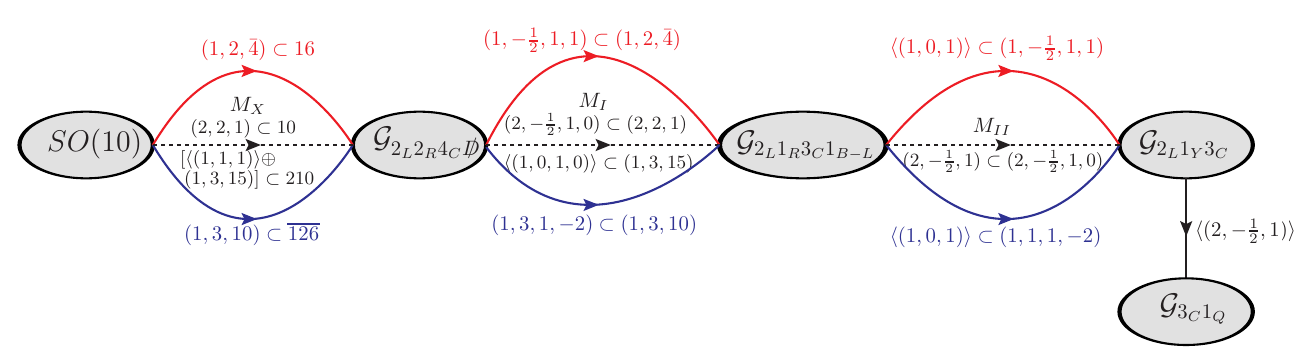}
	\end{center}
	\caption{$SO(10)$ breaking to the SM via two intermediate steps. The VEVs causing the successive symmetry breakings and the Higgs fields contributing to the RGEs at each stage are indicated. The upper (red) arrows correspond to the model 
	in Sec.~\ref{SO10_A}, and the lower (blue) arrows correspond to the model 
	in Sec.~\ref{SO10_B}. The dashed arrows in the middle are common to both models.}
	\label{fig:SO10_AB}
\end{figure}

The two symmetry breaking chains of $SO(10)$ considered in this paper are summarized in 
Fig.~\ref{fig:SO10_AB} together with the VEVs causing the breakings and the Higgs representations contributing to the RGEs at each stage. 
We first discuss the scenario where the $SO(10)$ is broken to 
$\mathcal{G}_{2_L2_R4_C}$ using the 210-plet VEV along its $(1,1,1)$ component, which also breaks $D$-parity 
\cite{Kibble:1982dd} that interchanges the representations of $SU(2)_L$ and $SU(2)_R$ and conjugates that of $SU(4)_C$. So the unbroken group is denoted as $\mathcal{G}_{2_L2_R4_C\slashed{D}}$. To break 
$SU(4)_C$ to $\mathcal{G}_{3_C1_{B-L}}$ and  
$SU(2)_R$ to $U(1)_R$, we employ the component $(1,3,15)\subset 210$, which produces $SU(4)_C$ and 
$SU(2)_R$ monopoles \cite{Lazarides:2019xai}. The breaking of $U(1)_{B-L} \otimes U(1)_R$ to $U(1)_Y$ is achieved by a VEV either along the component $(1,-\frac{1}{2},1,1)\subset (1,2,\bar{4})$ from a $16$-plet of $SO(10)$, or the component $(1,1,1,-2)\subset(1,3,10)$ from a $\overline{126}$-plet. These are two physically distinct cases \cite{Lazarides:2019xai}.
In the former case, the $SU(4)_C$ and $SU(2)_R$ monopoles, if not inflated away, eventually come together to form a double charged monopole and no cosmic strings are produced. In the latter case, however, in addition to the monopoles, we have necklaces with $SU(4)_C$ and $SU(2)_R$ monopoles and antimonopoles as well as stable $\mathbb{Z}_2$ cosmic strings. For a recent study with a single intermediate step unification with Abelian mixing see Refs.~\cite{Chakrabortty:2009xm,Chakrabortty:2017mgi,Ohlsson:2020rjc}.


In each of these cases the Abelian gauge coupling mixing is accounted for, as discussed in the previous section, by introducing a $2\times 2$ gauge coupling matrix in the $(R, B-L\equiv X)$ space:
\begin{equation}
G=
\begin{pmatrix}
g_{RR} & g_{RX} \\
g_{XR} & g_{XX}
\end{pmatrix}.
\end{equation}
At the breaking scales $M_I$ and $M_{II}$, the inverse squared gauge couplings 
$\omega(GG^{\rm T})^{-1}\omega^{\rm T}$ are suitably projected to match with the parent or daughter inverse squared gauge couplings respectively. At $M_{I}$, we use the projectors $\omega=(1,0)$ and $\omega=(0,1)$ to match the inverse squared gauge couplings with $1/g_{2R}^2$ and $1/g_{4C}^2$ respectively, and, at 
$M_{II}$, we take $\omega=(\sqrt{3/5},\sqrt{2/5})$ so that the inverse squared gauge couplings are projected on the hypercharge inverse squared gauge coupling. We consider the off-diagonal couplings 
\begin{equation}\label{g_and_r}
g_{RX}=g_{XR}=g \; \;  \mathrm{and} \; \; g_{XX}/g_{RR}=r 
\end{equation}
as free parameters in our subsequent analysis. Let us now discuss the two different breaking patterns of $SO(10)$ in turn.

\subsection{$SO(10)$ without Strings or Necklaces}\label{SO10_A}
In the $SO(10)$ case with no strings or necklaces, the breaking of $SO(10)$ to $\mathcal{G}_{2_L2_R4_C \slashed{D}}$ is achieved by the VEV of the component $(1,1,1) \subset210$. At this level, the component $(1,3,15) \subset 210$ remains massless. Also, the components $(2,2,1) \subset 10$ and $(1,2,\bar{4})\subset 16$. The next breaking to $\mathcal{G}_{2_L1_R3_C1_{B-L}}$ is induced by the VEV of $(1,3,15)\subset 210$, and we are left with a single massless electroweak Higgs doublet and a massless complex singlet $(1,-1/2,1,+1) \subset(1,2,\bar{4})$, whose VEV causes the subsequent breaking to the SM gauge symmetry.
The $SO(10)$ breaking chain considered here can be depicted as follows:
\begin{equation*}
SO(10)\xrightarrow[\langle 210 \rangle]{M_X}\mathcal{G}_{2_L 2_R 4_C\slashed{D}}\xrightarrow[\langle (1,3,15)\rangle\subset 210 ]{M_I} \mathcal{G}_{2_L 1_R 3_C1_{B-L}}\xrightarrow[\langle (1,-\frac{1}{2},1,1)\rangle\subset (1,2,\bar{4})\subset 16] {M_{II}}\mathcal{G}_{2_L 1_Y 3_C}.
\end{equation*}
 The Higgs fields which remain massless at each stage of the symmetry breaking and thus contribute to the RGEs are summarized in Table~\ref{tab1}. The $\beta$-coefficients and the RGEs are given in \hyperref[Appendix]{Appendix}.
\begin{table}[h!]
	\begin{center}
		\begin{tabular}{|c|c|c|c|}
			\hline
			$SO(10)$& $\mathcal{G}_{2_L 2_R 4_C\slashed{D}}$ & $\mathcal{G}_{2_L 1_R 3_C1_{B-L}}$ & $\mathcal{G}_{2_L 1_Y 3_C}$ \\
			\hline
			$10$ & $(2,2,1)$ & $(2,-{\frac{1}{2}},1,0)$ & $(2,-\frac{1}{2},1)$ \\
			$16$ & $(1,2,\bar{4})$& $(1,-\frac{1}{2},1,1)$ &  \\
			$210$ & $(1,3,15)$ &  & \\
			\hline
			\end{tabular}
		\caption{Higgs representations that contribute to the RGEs at each stage of the symmetry breaking for the $SO(10)$ model in Sec.~\ref{SO10_A}.}		\label{tab1}
	\end{center}

\end{table}

\subsection{$SO(10)$ with Strings and Necklaces}\label{SO10_B}

In the $SO(10)$ case with strings and necklaces, the breaking to $\mathcal{G}_{2_L2_R4_C \slashed{D}}$ is again achieved by the VEV of a scalar $210$-plet. At this level we again have the massless components $(1,3,15) \subset210$ and $(2,2,1) \subset10$, but now 
a massless $(1,3,10)\subset126$ too. The next breaking to $\mathcal{G}_{2_L1_R3_C1_{B-L}}$ is induced by 
$(1,3,15)$, and we are left with a massless $(1,1,1,-2) \subset(1,3,10)$ and a single electroweak Higgs doublet. The VEV of $(1,1,1,-2)$ does the breaking to the SM gauge group. The $SO(10)$ breaking chain considered here can be depicted as follows:
\begin{equation*}
SO(10)\xrightarrow[\langle 210 \rangle]{M_X}\mathcal{G}_{2_L 2_R 4_C\slashed{D}}\xrightarrow[\langle (1,3,15)\rangle\subset 210 ]{M_I} \mathcal{G}_{2_L 1_R 3_C1_{B-L}}\xrightarrow[\langle (1,1,1,-2)\rangle\subset (1,3,10)\subset \overline{126} ]{M_{II}}\mathcal{G}_{2_L 1_Y 3_C}.
\end{equation*}
The Higgs fields which contribute to the RGEs are summarized in Table~\ref{tab2}. The $\beta$-coefficients and the RGEs are given in \hyperref[Appendix]{Appendix}.
\begin{table}[h!]
	\begin{center}
		\begin{tabular}{|c|c|c|c|}
			\hline
			$SO(10)$& $\mathcal{G}_{2_L 2_R 4_C\slashed{D}}$ & $\mathcal{G}_{2_L 1_R 3_C1_{B-L}}$ & $\mathcal{G}_{2_L 1_Y 3_C}$ \\
			\hline
			$10$ & $(2,2,1)$ & $(2,-{\frac{1}{2}},1,0)$ & $(2,-\frac{1}{2},1)$ \\
			$\overline{126}$ & $(1,3,{10})$& $(1,1,1,-2)$ &  \\
			$210$ & $(1,3,15)$ &  & \\
			\hline
		\end{tabular}
		\caption{Higgs representations that contribute to the RGEs at each stage of the symmetry breaking for the $SO(10)$ model in Sec.~\ref{SO10_B}.} 	\label{tab2}
	\end{center}

\end{table}

\section*{$E_6$ Unification with Two Intermediate Steps}
\begin{figure}[htbp]
	\begin{center}
		\includegraphics[width=16cm]{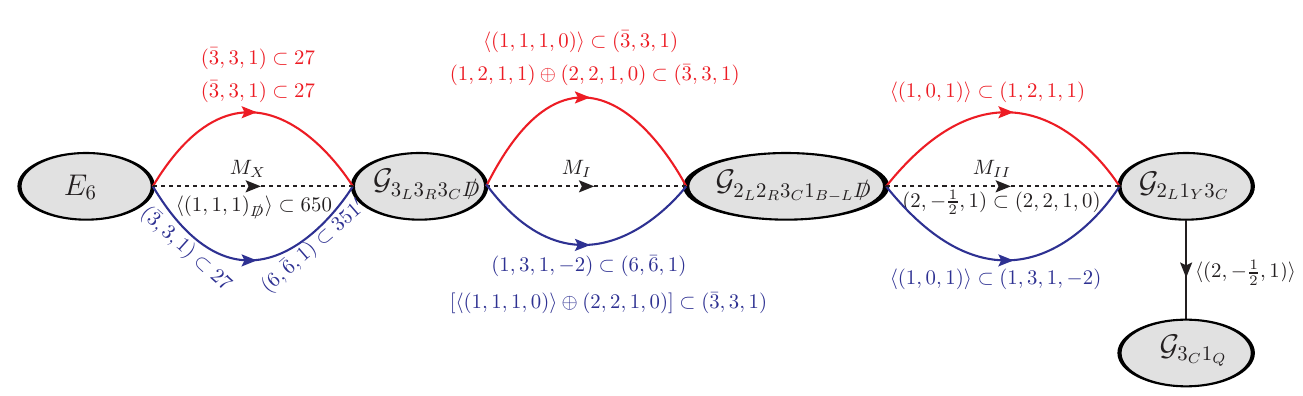}
	\end{center}
	\caption{$E_6$ breaking to the SM with two intermediate steps. The VEVs and the Higgs fields contributing to the RGEs at each stage are indicated. The upper (red) arrows correspond to the model 
	in Sec.~\ref{E6_C}, and the lower (blue) arrows correspond to the model 
	in Sec.~\ref{E6_D}. The dashed arrows in the middle are common to both models.}
	\label{fig:E6_CD}
\end{figure}
The two symmetry breaking patterns of $E_6$ discussed here are summarized in Fig.~\ref{fig:E6_CD} together with the VEVs causing the various symmetry breakings and the Higgs representations contributing to the RGEs at each stage. 
The $E_6$ gauge symmetry is broken to $\mathcal{G}_{3_L3_R3_C}$ using the $D$-violating VEV of a Higgs $650$-plet. The next breaking to $\mathcal{G}_{2_L2_R3_C1_{B-L}}$ is achieved using a $(\bar{3},3,1) \subset 27$ which, interestingly, is the $SO(10)$ singlet within the $27$-plet of $E_6$. The breaking to the SM gauge group though is induced through the VEV of a suitable sub-multiplet of a $27$ or alternatively of a $351^\prime$. From the perspective of emergence of possible topological defects, these are two distinct cases \cite{Lazarides:2019xai}. In the former case, i.e. with the Higgs $27$-plet, we have single and triply charged monopoles. But in the latter case, i.e. with a scalar $351^\prime$, we also have $\mathbb{Z}_2$ strings but not necklaces.  We now discuss these two cases
of $E_6$ breaking in turn.
\subsection{$E_6$ without Strings}\label{E6_C}
Let us first consider the $E_6$ case without strings. We use a Higgs $650$-plet to break $E_6$ to $\mathcal{G}_{3_L3_R3_C}$. At this level, we are left with two massless Higgs $(\bar{3},3,1)$ from the two $27$-plets. The breaking of $\mathcal{G}_{3_L3_R3_C}$ to $\mathcal{G}_{2_L2_R3_C1_{B-L}}$ is achieved through the VEV of the component $(1,1,1,0)$ in $(\bar{3},3,1)$. Noting the following decompositions of the $SU(3)_L$ and $SU(3)_R$ gauge bosons $(8,1,1)= (1,1,1,0)\oplus(3,1,1,0)\oplus(2,1,1,\pm 1)$, and $(1,8,1)= (1,1,1,0)\oplus(1,3,1,0)\oplus(1,2,1,\pm 1)$, it is evident that the nine would be Goldstone modes must transform as $(2,1,1,\pm 1)$, $(1,2,1,\pm 1)$, and one linear combination of two singlets. 
The components $(1,2,1,+1)$ and $(2,2,1,0)$ belonging to the  other $(\bar{3},3,1)$ remain massless. The breaking of $\mathcal{G}_{2_L2_R3_C1_{B-L}}$ to the SM gauge group is achieved by employing the $(1,2,1,+1)$ component. One linear combination of the $SU(2)_L$ doublets in the bi-doublet $(2,2,1,0)$ provides the electroweak Higgs doublet. The breaking chain of $E_6$ considered here can be depicted as follows:
\begin{equation*}
E_6\xrightarrow[\langle 650 \rangle]{M_X}\mathcal{G}_{3_L 3_R 3_C\slashed{D}}\xrightarrow[\langle (\bar{3},3,1)\rangle\subset 27]{M_I} \mathcal{G}_{2_L 2_R 3_C1_{B-L}\slashed{D}}\xrightarrow[\langle (1,2,1,1)\rangle\subset (\bar{3},3,1)\subset 27 ]{M_{II}} \mathcal{G}_{2_L 1_Y 3_C}.
\end{equation*}
 In Table~\ref{tab3}, we summarize the Higgs representations that contribute to the RGEs at each stage.
\begin{table}[h!]
	\begin{center}
		\begin{tabular}{|c|c|c|c|}
			\hline
			$E_6$ & $\mathcal{G}_{3_L 3_R 3_C\slashed{D}}$ & $\mathcal{G}_{2_L 2_R 3_C 1_{B-L}\slashed{D}}$ &  $\mathcal{G}_{2_L 1_Y 3_C}$ \\
			\hline
			$27$ & $(\bar{3},3,1)$ & $(2,2,1,0)$ & $(2,-\frac{1}{2},1)$ \\
			$27$ & $(\bar{3},3,1)$ & $(1,2,1,1)$ &  \\
			$650$ & & & \\
			\hline
		\end{tabular}
		\caption{Higgs representations that contribute to the RGEs at each stage of the symmetry breaking for the $E_6$ model in 
		Sec.~\ref{E6_C}.} 	\label{tab3}
	\end{center}

\end{table}

The relevant $\beta$-coefficients are given as:
\begin{align*} 
{\rm From} \ M_{II} \ {\rm to} \ M_{I} \ : & \;\;\;b_{2L} =-3, \; b_{2R} =-\frac{17}{6}, \; b_{3C} =-7, \; b_{B-L} =\frac{17}{4},  \;\;\; b_{ij}= 
\begin{pmatrix}
8 & 3 & 12 & \frac{3}{2} \\ 
3 & \frac{61}{6} & 12 & \frac{9}{4} \\ 
\frac{9}{2} & \frac{9}{2} & -26 & \frac{1}{2} \\ 
\frac{9}{2} & \frac{27}{4} & 4 & \frac{37}{8}
\end{pmatrix}. \\ 
{\rm From} \ M_{I} \ {\rm to} \ M_X \ : & \;\;\;b_{3L} =-4, \; b_{3R} =-4, \; b_{3C} =-5,  \;\;\; b_{ij}= 
\begin{pmatrix}
34 & 28 & 12 \\
 28 & 34 & 12 \\
 12 & 12 & 12
\end{pmatrix}.
\end{align*}

\subsection{$E_6$ with Strings}\label{E6_D}
In order to have cosmic strings, we break $E_6$ to $\mathcal{G}_{3_L3_R3_C \slashed{D}}$ by again employing a scalar $650$-plet, but at this level we are left with the following Higgs massless modes: $(\bar{3},3,1)\subset 27$ and $(6,\bar{6},1) \subset{351'}$. The breaking of $\mathcal{G}_{3_L3_R3_C \slashed{D}}$ to $\mathcal{G}_{2_L2_R3_C1_{B-L} \slashed{D}}$ is again achieved by the VEV of 
$(1,1,1,0)\subset(\bar{3},3,1)$, and we are left with two massless Higgs representations, namely $(2,2,1,0)\subset(\bar{3},3,1)$ and 
$(1,3,1,-2)\subset(6,\bar{6},1)$. The latter component breaks $\mathcal{G}_{2_L2_R3_C1_{B-L} \slashed{D}}$ to the SM gauge group and we end up with a massless SM Higgs doublet from the former sub-multiplet, as in the previous cases. The breaking chain of
$E_6$ considered here can be depicted as follows:
\begin{equation*}
E_6\xrightarrow[\langle 650 \rangle]{M_X}\mathcal{G}_{3_L 3_R 3_C\slashed{D}}\xrightarrow[\langle (\bar{3},3,1)\rangle\subset 27]{M_I} \mathcal{G}_{2_L 2_R 3_C1_{B-L}\slashed{D}}\xrightarrow[\langle (1,3,1,-2)\rangle\subset (6,\bar{6},1)\subset 351']{M_{II}}\mathcal{G}_{2_L 1_Y 3_C}.
\end{equation*}
The Higgs fields contributing to the RGEs at each stage are presented in Table~\ref{tab4}
\begin{table}[h!]
	\begin{center}
		\begin{tabular}{|c|c|c|c|}
			\hline
			$E_6$ & $\mathcal{G}_{3_L 3_R 3_C\slashed{D}}$ & $\mathcal{G}_{2_L 2_R 3_C 1_{B-L}\slashed{D}}$ &  $\mathcal{G}_{2_L 1_Y 3_C}$ \\
			\hline
			$27$ & $(\bar{3},3,1)$ & $(2,2,1,0)$ & $(2,-\frac{1}{2},1)$ \\
			$351^\prime$ & $(6,\bar{6},1)$& $(1,3,1,-2)$ &  \\
			$650$ & & & \\
			\hline
		\end{tabular}
		\caption{Higgs representations that contribute to the RGEs at each stage of the symmetry breaking for the $E_6$ model in Sec.~\ref{E6_D}.}	\label{tab4}
	\end{center}

\end{table}

The necessary $\beta$-coefficients of the relevant RGEs are:
\begin{align*} 
{\rm From} \ M_{II} \ {\rm to} \ M_{I} \ : & \;\;\;b_{2L} =-3, \; b_{2R} =-\frac{7}{3}, \; b_{3C} =-7, \; b_{B-L} =\frac{11}{2},  \;\;\; b_{ij}= 
\begin{pmatrix}
8 & 3 & 12 & \frac{3}{2} \\ 
3 & \frac{80}{3} & 12 & \frac{27}{2} \\ 
\frac{9}{2} & \frac{9}{2} & -26 & \frac{1}{2} \\ 
\frac{9}{2} & \frac{81}{2} & 4 & \frac{61}{2}
\end{pmatrix}. \\ 
{\rm From} \ M_{I} \ {\rm to} \ M_X \ : & \;\;\;b_{3L} =\frac{1}{2}, \; b_{3R} =\frac{1}{2}, \; b_{3C} =-5,  \;\;\; b_{ij}= 
\begin{pmatrix}
253 & 220 & 12 \\ 
220 & 253 & 12 \\ 
12 & 12 & 12
\end{pmatrix}.
\end{align*}

\section{Unification with Threshold Corrections and Proton Decay}
\label{sec:unification}

We aim to find the unification solutions in terms of the unified gauge coupling constant $g_U$, the unification scale $M_X$, and the intermediate scales $M_{I,II}$ that are consistent with the experimental observables at the $Z$ gauge boson mass $m_Z$. To perform this task, we define a $\chi^2$ statistic at $m_Z$ as 
\begin{equation}
\chi^2 = {\sum_{i=1}^3 \frac{\left(g_i^2 - g_{i,\mathrm{exp}}^2\right)^2}{\sigma^2_{g^2_{i,\mathrm{exp}}}}},
\end{equation}
which we minimize to find the unification solutions. Here, $g_i$ ($i=Y, 2L, 3C$) are the SM gauge couplings at $m_Z$ and are related to the unification and intermediate scales and the unified gauge coupling through the RGEs. On the other hand, $g_{i,\mathrm{exp}}^2$ are their experimental values squared computed from the electroweak observables along with the standard deviations denoted by 
$\sigma$ -- see Table~\ref{ewpo}.
\begin{table}[htb!]
 	\begin{center}
 		\begin{tabular}{|c|c|}
 			\hline
 			 $Z$-boson mass $m_Z$ & $91.1876(21)$ GeV \\
 			\hline
 			Strong fine structure constant $\alpha_{3C}$ & $0.1185(6)$ \\
 			\hline
 			Fermi coupling constant $G_F$ & $1.1663787(6)\times 10^{-5} \ \rm{GeV}^{-2}$ \\
 			\hline
 			Weinberg angle $\sin^2{\theta_W}$ & $0.23126(5)$ \\
 			\hline
 		\end{tabular}
 		\caption{Experimental observables at $m_Z$.}\label{ewpo}
 	\end{center}
 \end{table}
This method ensures that our unification solutions are consistent with the electroweak observables \cite{Tanabashi:2018oca}. We have taken the solutions for which the $\chi^2_{min}<1$.

In the $E_6$ case, we add suitable threshold corrections while implementing the matching conditions at the breaking scales. Without loss of generality, we assume that the ratio of the mass of the heavy fields belonging to the parent symmetry to the symmetry breaking scale $\mu$, i.e. $M_i/\mu$ ($i=V,S,F$ with notation as in Eq.~(\ref{lamda})), varies within the range $[1/2,2]$. In the case of the $SO(10)$ breaking chains, the presence of the Abelian mixing leads to a range of allowed solutions. Thus, additional contributions due to threshold corrections can be ignored to reduce the number of free parameters. In passing, we would like to mention that inclusion of threshold corrections will only widen the allowed parameter space without invalidating our conclusion.

One of the most interesting predictions of GUTs is the possibility of proton decay, which is unfortunately yet to be observed. In the ongoing experiments, the proton lifetime is continuously pushed to larger values that, in turn, puts severe constraints on the unification scale. Our aim is to find the unification solutions that are simultaneously compatible with the low energy observables and  the exclusion limits on proton lifetime. Here, we consider the decay of proton into a positron and a neutral pion. The partial lifetime for this channel is given as \cite{Weinberg:1979sa, Wilczek:1979hc, Weinberg:1980bf, Abbott:1980zj,Lucha:1984tv,FileviezPerez:2004hn, Nath:2006ut}
\begin{eqnarray}
\tau_p &= & \Bigg[ \frac{m_p}{32\pi}\left(1-\frac{m_{\pi^0}^2}{m_p^2}\right)^2 A_L^2 \frac{g_U^4}{4 M_X^4}(1+|V_{ud}|^2)^2  \nonumber \\
& & \times  \left( A_{SR}^2 |\langle \pi^0 \rvert (ud)_R u_L\lvert p \rangle |^2 + A_{SL}^2 |\langle \pi^0 \rvert (ud)_L u_L\lvert p \rangle |^2 \right) \Bigg]^{-1},
\end{eqnarray}
where $g_U$ is the unified gauge coupling, and $m_p$ and $m_{\pi^0}$ are the masses of the proton and neutral pion respectively. The coefficients 
$A_{SR(SL)}$ include the enhancement factors due to the RGEs for proton decay operators from  $M_X$ to $m_Z$ \cite{Buras:1977yy,Abbott:1980zj,Goldman:1980ah,Caswell:1982fx,DANIEL1983219,Ibanez:1984ni,MUNOZ198655}, and $A_L$ denotes the renormalization factor from $m_Z$ to the QCD scale ($\sim 1$ GeV) \cite{Nihei:1994tx}. The 
Cabibbo–Kobayashi–Maskawa matrix element $V_{ud}$ is given by $|V_{ud}| = 0.9742$ \cite{Tanabashi:2018oca} and the form factors are taken from the lattice QCD computation of Ref.~\cite{Aoki:2017puj}:
\begin{equation}
\langle \pi^0 \rvert (ud)_R u_L\lvert p \rangle  = -0.131, \ \ \langle \pi^0 \rvert (ud)_L u_L\lvert p \rangle = 0.134 \ .
\end{equation}

\begin{figure}[htbp]
 	\begin{center}
 		\subfloat[Contour Plot of $M_X$.]
 		{
 			\includegraphics[scale=0.30]{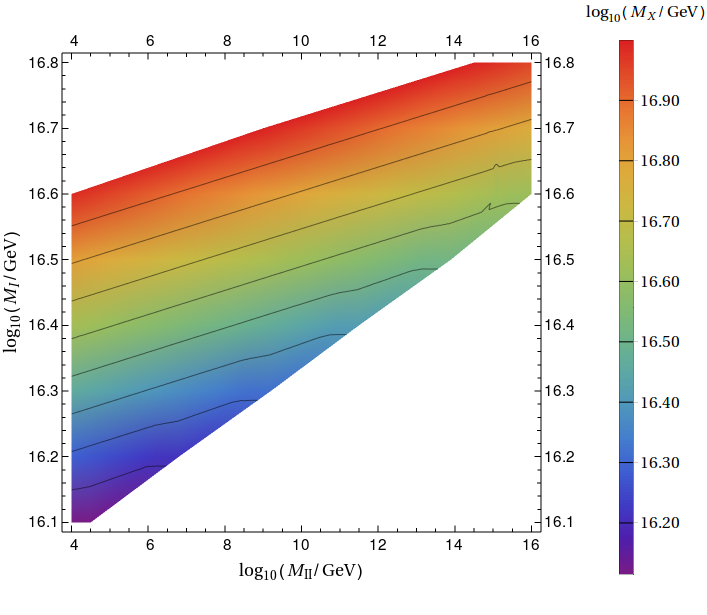}
 		}
 		\subfloat[Contour Plot of $\tau_p$.]
 		{
 			\includegraphics[scale=0.30]{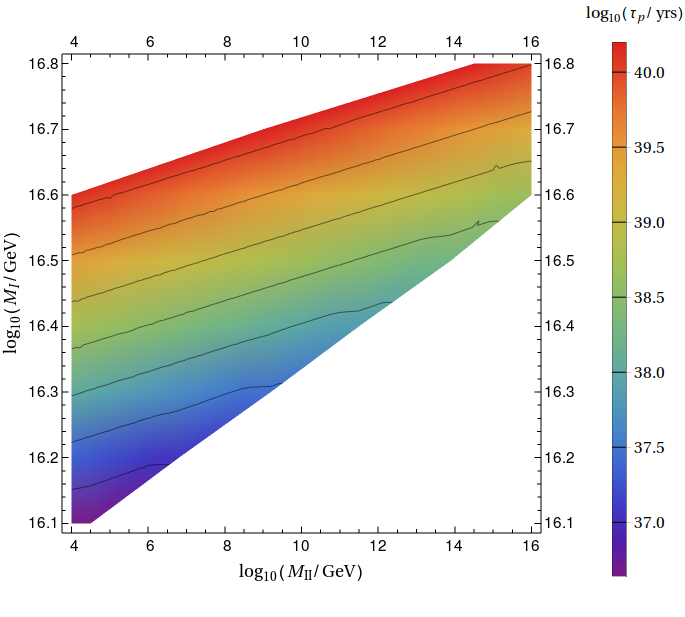}
 		}
 		\caption{Contour plots for the $SO(10)$ breaking chain in Sec.~\ref{SO10_A}, where the symmetry breaking 
		$\mathcal{G}_{2_L1_R3_C1_{B-L}}\to \mathcal{G}_{2_L1_Y3_C}$ at $M_{II}$ is achieved by the VEV of 
		$(1,-1/2,1,1)\subset 16$. For this fit, we have $g_U \in [0.52, 0.53]$, $r\in [0,1]$, and $g\in [0.40,0.60]$.}
		\label{fig:so10_16}
 	\end{center}
 \end{figure}

\begin{figure}[htbp]
 	\begin{center}
 		\subfloat[Contour Plot of $M_X$.]
 		{
 			\includegraphics[scale=0.30]{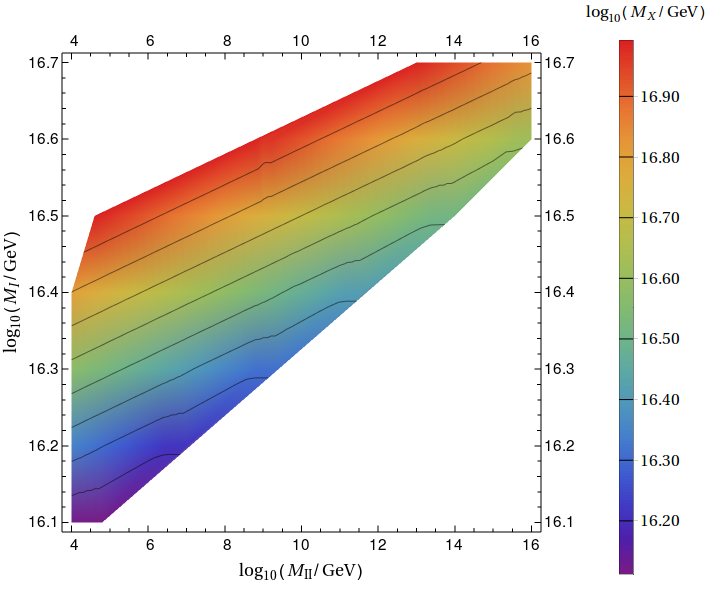}
 		}
 		\subfloat[Contour Plot of $\tau_p$.]
 		{
 			\includegraphics[scale=0.30]{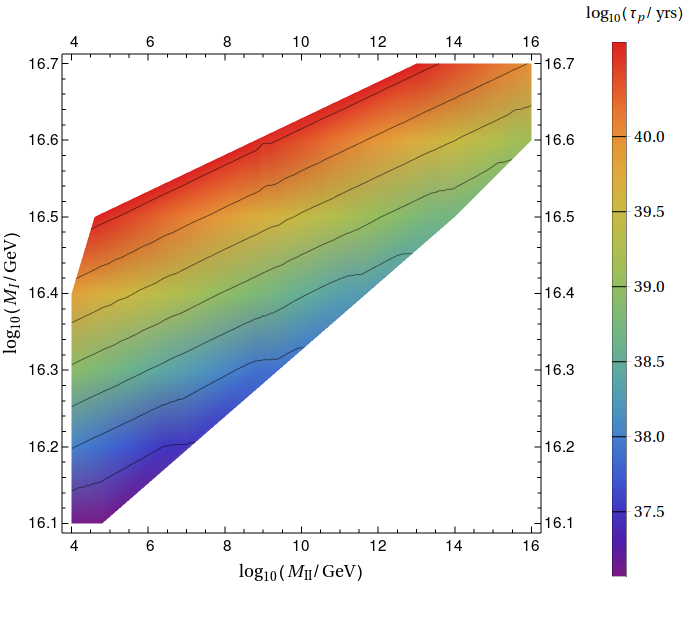}
 		}
 		\caption{Contour plots for the $SO(10)$ breaking chain in Sec.~\ref{SO10_B}, where the VEV of 
 			$(1,1,1,-2)\subset\overline{126}$ breaks 
 			$\mathcal{G}_{2_L1_R3_C1_{B-L}}$ to $\mathcal{G}_{2_L1_Y3_C}$. For this fit, we have 
 			$g_U \in [0.52, 0.53]$, $r\in [0,1]$, and $g\in [0.39,0.59]$.}
 		\label{fig:so10_126}
 	\end{center}
 \end{figure}

\begin{figure}[htbp]
\begin{center}
\subfloat[Contour Plot of $M_X$.]
{
\includegraphics[scale=0.30]{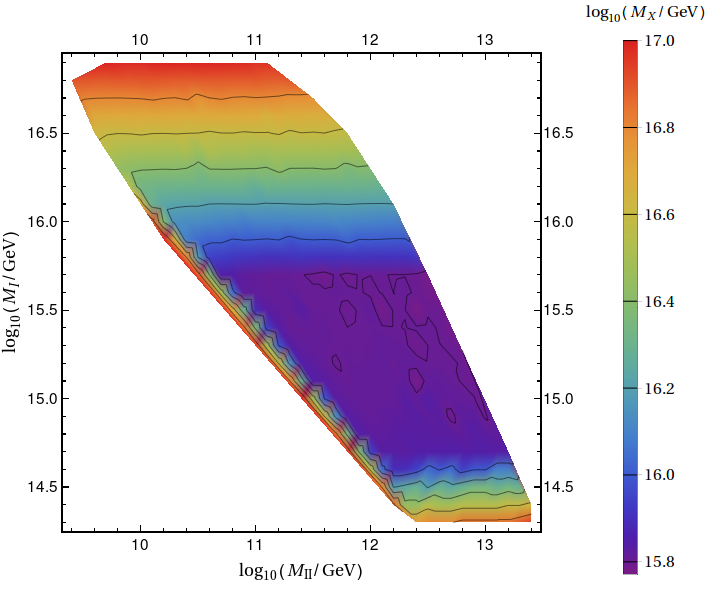}
}
\subfloat[Contour Plot of $\tau_p$.]
{
\includegraphics[scale=0.30]{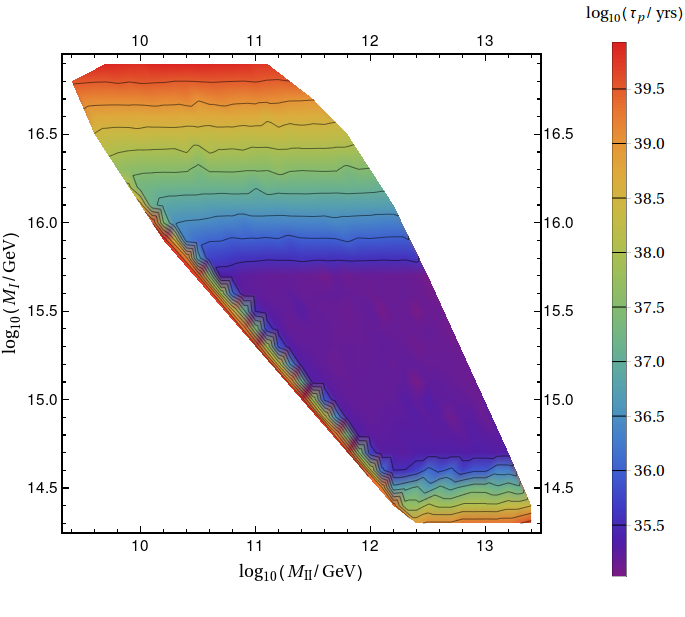}
}
\caption{Contour plots for the $E_6$ breaking chain in Sec.~\ref{E6_C}, where the VEV of $(1,2,1,1)\subset 27$ causes the symmetry breaking $\mathcal{G}_{2_L2_R3_C1_{B-L}}\to \mathcal{G}_{2_L1_Y3_C}$ at $M_{II}$. For this fit, $g_U \in [0.51, 0.54]$.}
\label{fig:e6_27}
\end{center}
\end{figure}

\begin{figure}[htbp]
\begin{center}
\subfloat[Contour Plot of $M_X$.]
{
\includegraphics[scale=0.30]{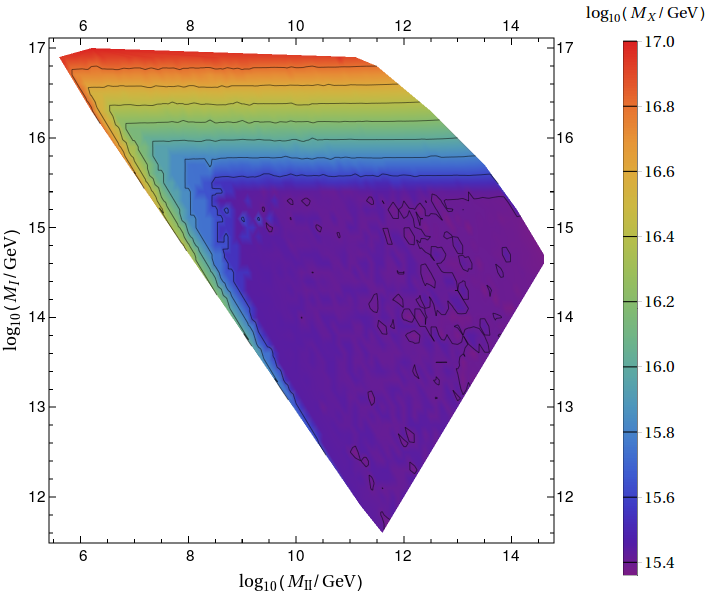}
}
\subfloat[Contour Plot of $\tau_p$.]
{
\includegraphics[scale=0.30]{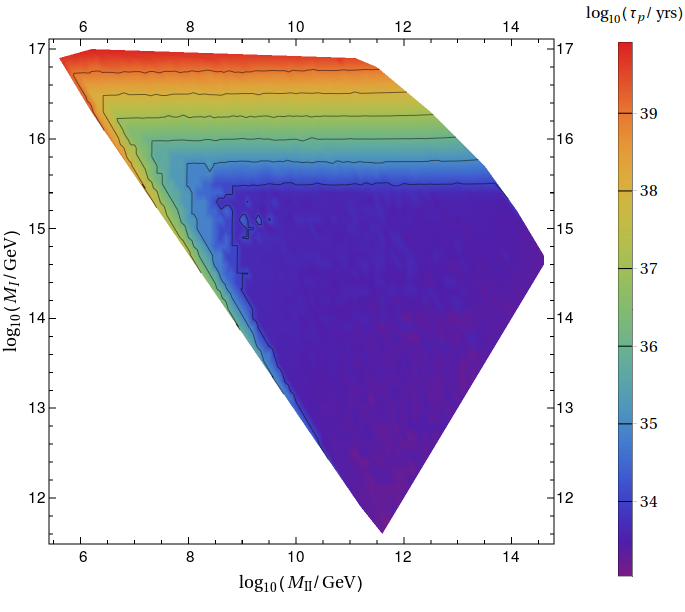}
}
\caption{Contour plots for the $E_6$ breaking chain in Sec.~\ref{E6_D}, where the VEV of $(1,3,1,-2)\subset 351'$ causes the symmetry breaking $\mathcal{G}_{2_L2_R3_C1_{B-L}}\to \mathcal{G}_{2_L1_Y3_C}$ at $M_{II}$. For this fit, 
$g_U \in [0.51, 0.56]$.} \label{fig:e6_351}
\end{center}
\end{figure}

We construct the unification solutions with unification scale up to $M_X = 10^{17}$ GeV which are consistent with all the constraints mentioned above for each of the symmetry breaking chains in Secs.~\ref{SO10_A}, \ref{SO10_B}, \ref{E6_C}, and \ref{E6_D} and depict them in Figs.~\ref{fig:so10_16}, \ref{fig:so10_126}, \ref{fig:e6_27}, and \ref{fig:e6_351} respectively. In particular, we present the allowed values of the unification scale $M_X$ and the partial proton lifetime $\tau_p$ as functions of $M_I$ and $M_{II}$. We note that the unified gauge coupling $g_U$ lies within the range $[0.52, 0.53]$ for the two $SO(10)$ breaking chains in Secs.~\ref{SO10_A} and \ref{SO10_B}. In the case of $E_6$, $g_U$ ranges within $[0.51,0.54]$ and $[0.51,0.56]$ for the models in Secs.~\ref{E6_C} and \ref{E6_D} respectively. In $SO(10)$, where we have Abelian mixing at the second intermediate symmetry breaking, we find unification solutions for $g$ within $[0.40,0.60]$ and $[0.39,0.59]$ for the models is Secs.~\ref{SO10_A} and \ref{SO10_B} respectively, and $r\in [0,1]$ for both cases -- for the definition of $g$ and $r$ see 
Eq.~(\ref{g_and_r}). For the breaking chain in Sec.~\ref{SO10_A}, we find that the intermediate scales lie in the ranges 
$\log_{10} (M_I/\mathrm{GeV})$ $\in [16.1,16.8]$ and $\log_{10} (M_{II}/\mathrm{GeV})$ $\in [4.0,16.0]$, and for the chain in Sec.~\ref{SO10_B} in the ranges $\log_{10} (M_I/\mathrm{GeV})$ $\in [16.1,16.7]$ and $\log_{10} (M_{II}/\mathrm{GeV})$ 
$\in [4.0,16.0]$. For $E_6$ , we obtain the intermediate scales for the model in Sec.~\ref{E6_C} in the ranges 
$\log_{10} (M_I/\mathrm{GeV})$ $\in [14.3,16.9]$ and $\log_{10} (M_{II}/\mathrm{GeV})$ $\in [9.4,13.4]$, and for the chain in 
Sec.~\ref{E6_D} in the ranges $\log_{10} (M_I/\mathrm{GeV})$ $\in [11.6,17.0]$ and $\log_{10} (M_{II}/\mathrm{GeV})$ 
$\in [5.6,14.6]$. At this point, we have verified that all the unification solutions satisfy the present Super-Kamiokande limit ($\tau_p > 1.6\times 10^{34}$ years)\cite{Miura:2016krn,Takhistov:2016eqm}, and also the projected Hyper-Kamiokande limit 
($\tau_p > 8.0\times 10^{34}$ years) \cite{Yokoyama:2017mnt}. 
 In the next section, we will find the ranges of $M_X$, $M_I$, and $M_{II}$ for which a successful GUT-inflation scenario with a Coleman-Weinberg potential is compatible with the Planck satellite results \cite{Akrami:2018odb}.
 
\section{Inflation with Coleman-Weinberg Potential}
\label{sec:inflation}
In order to understand the inflationary dynamics, we consider the relevant part of the scalar potential \cite{Shafi:1983bd,Lazarides:1984pq,Shafi:2006cs}
\begin{equation}
V = \frac{\lambda}{4}\phi^4 - \frac{1}{2}\beta^2\phi^2\chi^2 + \frac{a}{4}\chi^4 + A\phi^4\left[ \log\left(\frac{\phi}{M}\right)+c\right] + V_0 \ ,
\label{eq:pot}
\end{equation}
where the GUT-singlet inflaton field $\phi$ and  the GUT symmetry breaking scalar  $\chi$ are  canonically normalized real scalar fields, and $A=\beta^4\, D/16\pi^2$ \cite{Lazarides:2019xai}, with $D$ being the dimensionality of the representation to which $\chi$ belongs. We substitute $\chi=(\beta/\sqrt{a})\phi$ in Eq.~(\ref{eq:pot}), which minimizes the potential for any given value of $\phi$. In the limit $\lambda << \beta^4$ and requiring that the potential is minimized at 
$\phi=M$ with $V(\phi=M)=0$, we find
\begin{equation}\label{CW-potential}
V(\phi)= A\phi^4\left[ \log\left(\frac{\phi}{M}\right) - \frac{1}{4}\right] +\frac{AM^4}{4},
\end{equation}
where $V_0=AM^4/4$.

The slow-roll parameters can be written in terms of the potential and its derivatives as follows (for a review see Ref.~\cite{Lyth:2009zz}):
\begin{equation}\label{slow-roll-params}
\epsilon = \frac{m_{\rm Pl}^2}{2}\left(\frac{V^\prime}{V} \right)^2 \ , \; \; \eta =m_{\rm Pl}^2 \frac{V^{\prime\prime}}{V} \ , \; \; \xi^2 = m_{\rm Pl}^4 \frac{V^\prime V^{\prime\prime\prime}}{V^2},
\end{equation}
where $m_{\rm Pl}$ is the reduced Planck scale and primes represent derivatives with respect to $\phi$.
The spectral index $n_s$, the tensor-to-scalar ratio $r$, and the running of the 
spectral index $\alpha\equiv d{ n_s}/d\,{\mathrm{ln} k}$ can be deduced using the slow-roll parameters computed at the pivot scale 
$k_*$:
\begin{equation}\label{ns-r-alpha}
n_s = 1-6\epsilon_* + 2\eta_* \ , \; \; r = 16\epsilon_* \ , \; \; \alpha =16\epsilon_*\eta_* -24 \epsilon_*^2 -2 \xi_*^2.
\end{equation}
Here, the subscript $*$ signifies the values of the parameters at the pivot scale $k_*=0.05 \ \mathrm{Mpc}^{-1}$. The experimental values of these observables at $95\%$ confidence level are as follows \cite{Akrami:2018odb}: 
\begin{equation}\label{infl_obs}
n_s = 0.9658 \pm 0.0080,  \
r < 0.068, \ \mathrm{and} \ \alpha = -0.0066 \pm 0.0140 .
\end{equation}
The amplitude of the curvature perturbation $\Delta_R$ is given by
\begin{equation}\label{curv_per_amp}
\Delta_R^2=\frac{1}{12\pi^2 m_{\rm Pl}^6}\frac{V^3}{(V^\prime)^2}\Big|_{\phi=\phi_*} \ ,
\end{equation}
with its experimental value ${\Delta_R^2}^{exp} = (2.099\pm 0.101)\times 10^{-9}$ at 
$95\%$ confidence level \cite{Akrami:2018odb}.

The number of $e$-foldings for the pivot scale is computed using the following equation:
\begin{equation}\label{N_1}
N_* =\frac{1}{m_{\mathrm{Pl}}^2} \int_{\phi_e}^{\phi_*} \frac{V \mathrm{d}\phi}{V^\prime} \ .
\end{equation}
Here $\phi_e$ is the value of $\phi$ at the end of inflation and is deduced using the following condition:
\begin{equation}
max(|\eta|,\epsilon)=1 \ .
\end{equation}
The number of $e$-foldings for the pivot scale $k_*=0.05 \ \mathrm{Mpc^{-1}}$ can alternatively be obtained from the knowledge of the thermal history of the universe \cite{Liddle:2003as}: 
\begin{equation}\label{N_2}
N_* \simeq 61.5 + \frac{1}{2} \mathrm{ln} \frac{\rho_*}{m_{\rm Pl}^4}-\frac{1}{3(1+\omega_r)} \mathrm{ln} \frac{\rho_e}{m_{\rm Pl}^4} + \left(\frac{1}{3(1+\omega_r)} - \frac{1}{4} \right)\mathrm{ln} \frac{\rho_r}{m_{\rm Pl}^4} \ ,
\end{equation}
where $\rho_* = V(\phi_*)$, $\rho_e = V(\phi_e)$, and $\rho_r = (\pi^2/30) g_*T_r^4$ are the energy densities at the pivot scale, at the end of inflation, and at the reheat temperature $T_r$, respectively, and $\omega_r$ is the effective equation-of-state parameter from the end of inflation until reheating. The effective number of massless degrees of freedom $g_*$ at reheating is taken to be 106.75 corresponding to the SM spectrum. Clearly, the values of $N_*$ from
Eqs.~(\ref{N_1}) and (\ref{N_2}) must coincide. For our analysis, we consider the so-called 
middle-$N$ scenario (see Ref.~\cite{Senoguz:2015lba}), where $T_r = 10^9$ GeV and $\omega_r = 0$.

In order to find consistent inflationary solutions in terms of the  parameters $ A$, $M$, $\phi_*$, 
 and $\phi_e$, we construct a $\chi^2$ function and adopt the following steps:
\begin{itemize}
	
\item[(i)] We write
\begin{equation}\label{A_param}
A=4V_0/M^4, \quad M_X=\sqrt{8\pi/a} (V_0/D)^{1/4},
\end{equation}
and express $\epsilon$ as a function of $M,\; {\rm and}\; \phi$. The value $\phi_e$ of $\phi$ at the end of inflation is then determined by requiring that $\epsilon(M,\phi_e)=1$.

\item[(ii)] We express $n_s$ as a function of $M,\;\phi_*$, and $\Delta_R^2$ as a function of $V_0,\; M,
\; \phi_*$.

\item[(iii)] We compute $N_*$ as a function of $M$, $\phi_*$, $\phi_e $ from Eq.~(\ref{N_1}) and as a function 
of $V_0$, $M$, $\phi_*$, $\phi_e $ using Eq.~(\ref{N_2}). We then require that both results coincide up to a numerical tolerance $\Delta N_{*}$. 

\item[(iv)] We choose the following values for the observables:
\begin{eqnarray} \label{eq:observables}
(a) & {\Delta_R^2}^{exp} \pm \delta({\Delta_R^2}^{exp}) &= (2.099\pm 0.101)\times 10^{-9}, \nonumber \\
(b)& \epsilon(\phi_e)\pm\delta(\epsilon(\phi_e)) & = 1.0 \pm 0.1, \ \nonumber \\
(c)& \Delta N_{*} \pm \delta(\Delta N_{*} ) &= 0.0 \pm 1.0,  
\end{eqnarray}
for given values of $V_0$ and the other parameters $M$, $\phi_*$, and $\phi_e$. We ensure that the arbitrary choice of  tolerance for 
$\epsilon(\phi_e)$ and $\Delta N_{*} $ does not affect our conclusions.

\item[(v)] We define $\chi^2$ as a function of $M$, $\phi_* $, and $\phi_e$ for some benchmark choices of  $V_0$:
\begin{equation}
\chi^2 = \frac{\left({\Delta_R^2}- {\Delta_R^2}^{exp} \right)^2}{(\delta({\Delta_R^2}^{exp}) )^2} + \frac{(\epsilon(\phi_e)-1.0)^2}{(\delta \epsilon(\phi_e))^2} + \frac{(\Delta N_{*})^2}{(\delta(\Delta N_{*} ))^2}.
\end{equation}
We minimize the $\chi^2$ function to find the best fit values of $M$, $\phi_*$, and $\phi_e$ for a specific choice of  $V_0^{1/4}$. In the process of minimization, we also ensure that $\epsilon$ converges to unity before $|\eta|$ as $\phi$ approaches $\phi_e$.

\item[(vi)] Finally, using the best-fit values of the parameters, we estimate $A$ from Eq.~(\ref{A_param}) and, subsequently, reconstruct the potential using Eq.~(\ref{CW-potential}). We further compute the slow-roll parameters from Eq.~(\ref{slow-roll-params}) and also $n_s$, $r$, and $\alpha$ from Eq.~(\ref{ns-r-alpha}). 
\end{itemize}


\begin{table}[htbp]

\begin{center}
		\addtolength{\tabcolsep}{-2pt}
		\small
\begin{tabular}{| c | c | c | c | c | c | c | c | c | c | c |}
\hline
 $\frac{V_0^{1/4}}{10^{16}\text{GeV}}$ & $\frac{V\left(\phi_*\right)^{1/4}}{10^{16}\text{GeV}}$ & $\log_{10}A$ & ${M}/m_{\rm Pl}$ & $\phi_*/m_{\rm Pl}$ & $\phi_e/m_{\rm Pl}$ & $N_*$ & $\Delta_R^2\times 10^9$ & $n_s$ & $r$ & $\alpha \times 10^4$ \\
 \hline
 1.51 & 1.44 & -13.4 & 20.20 & 9.13 & 18.87 & 52.3 & 2.1 & 0.9584 & 0.039 & -6.41 \\
 1.59 & 1.50 & -13.5 & 21.89 & 10.52 & 20.55 & 52.3 & 2.1 & 0.9596 & 0.045 & -6.40 \\
 1.66 & 1.55 & -13.6 & 23.81 & 12.17 & 22.47 & 52.4 & 2.1 & 0.9606 & 0.052 & -6.41 \\
 1.74 & 1.59 & -13.6 & 26.01 & 14.09 & 24.65 & 52.4 & 2.1 & 0.9615 & 0.058 & -6.44 \\
 1.82 & 1.64 & -13.7 & 28.50 & 16.33 & 27.15 & 52.5 & 2.1 & 0.9623 & 0.065 & -6.49 \\
 \hline
\end{tabular}
\caption{Values of the parameters for successful inflation with a Coleman-Weinberg potential.}
\label{tab:Infl_para}
\end{center}
\end{table}
 In Table~\ref{tab:Infl_para}, we present the estimated values of the various parameters of the model including the slow-roll parameters, which are within two standard deviations from their central experimental values -- see Eqs.~(\ref{infl_obs}) and 
(\ref{eq:observables}). The range of the corresponding $V_0$ is $V_0^{1/4}/10^{16}~{\rm GeV} \in [1.51,1.82]$. We see that these solutions are perfectly compatible with all the requirements for a successful inflation. At this point it is worth mentioning that the minimum values of $\chi^2$ are found to be $\chi^2_{min}\sim 10^{-13}<<1$ for the fitted parameters in Table~\ref{tab:Infl_para}. Recall that the first step of symmetry breaking of $SO(10)$ and $E_6$ is achieved by the $210$- and $650$-dimensional representation, respectively. Therefore, using Eq.~(\ref{A_param}), we find that the unification scale $M_X$ is given by $M_X = 2.342 \ V_0^{1/4}$ for $SO(10)$ and 
$M_X = 1.766 \ V_0^{1/4}$ for $E_6$. Thus, the range of the unification scale for successful inflation is $\log_{10}(M_X/\mathrm{GeV})\in [16.55, 16.63]$ and $\log_{10}(M_X/\mathrm{GeV})\in [16.43, 16.51]$ for $SO(10)$ and $E_6$, respectively, which are compatible with the present Super-Kamiokande \cite{Miura:2016krn} and future Hyper-Kamiokande \cite{Yokoyama:2017mnt} bounds on proton lifetime.

Before concluding this section let us emphasize that the tensor-to-scalar ratio $r$ is predicted to lie somewhere around $0.03-0.06$ -- see Table~\ref{tab:Infl_para}. In the presence of  non-minimal coupling to gravity, $r$ can approach values close to 0.003 \cite{Okada:2010jf,Bostan:2018evz}.

\section{Phase Transitions and Formation of Topological Defects}
\label{sec:phasetrans}
The first step of the spontaneous breaking of $E_6$ and $SO(10)$ is achieved through the VEV of a suitable GUT non-singlet canonically normalized real scalar field $\chi$, which sets the value of the unification scale $\langle\chi\rangle\equiv M_X=(\beta/\sqrt{a}) M$. We chose $\chi$ to belong to a $650$- or $210$-plet of $E_6$ or $SO(10)$, respectively.
At this point a legitimate question to ask is when the actual GUT phase transition takes place. In the absence of temperature corrections, the potential is minimized at $\langle\chi\rangle= \pm (\beta/\sqrt{a})\phi$ for any given value of $\phi$. Thus, the field $\chi$ remains non-zero as $\phi$ rolls towards $M$ from non-zero values. However, during inflation, we must include in the potential the temperature correction $(1/2)\sigma_\chi T_H^2 \chi^2$ \cite{Shafi:1983bd}, where $T_H={H}/{2\pi}$ is the Hawking temperature and $\sigma_\chi$ is assumed to be of order unity. Initially, $\phi$ is small and this correction term dominates over the second term in the right hand side of Eq.~(\ref{eq:pot}). Consequently, the potential attains its minimum at $\chi=0$ with the GUT gauge symmetry restored. But as $\phi$ grows, $\chi=0$ turns into a local maximum of the potential and two global minima appear at 
\begin{equation}
\chi=\pm \sqrt{[\beta^2\phi^2-\sigma_\chi T_H^2]/a} \ .
\end{equation}
The potential difference between the local maximum at $\chi=0$ and these minima is
\begin{equation}
\Delta V=(\beta^2\phi^2-\sigma_\chi T_H^2)^2/4a.
\end{equation} 

At the beginning, these minima are very shallow and the fluctuations between them over the local maximum are very frequent. The fluctuations occur within spheres of radii equal to the Higgs correlation length $m_{\mathrm{eff}}^{-1}$, with $m_\mathrm{eff}$ being the effective mass of $\chi$ at the minima given by 
\begin{equation}
\label{effective-mass_infl}
m_{\mathrm{eff}}^2=2[\beta^2\phi^2-\sigma_\chi T_H^2].
\end{equation}
These fluctuations are Boltzmann suppressed when the  energy required is higher than $T_H$. This gives the so-called Ginzburg criterion \cite{GINZBURG}:
\begin{equation} \label{form_mon}
\frac{4\pi}{3}m_{\mathrm{eff}}^{-3}\Delta V > T_H \Rightarrow \beta^2\phi^2 > \left( \frac{72a^2}{\pi^2}+\sigma_\chi\right) T_H^2 .
\end{equation}
At the value of $\phi$ saturating this inequality, $\chi$ settles down in one of the vacua and the breaking of the GUT gauge symmetry is completed leading to the formation of topological defects.

The next (first intermediate) step of symmetry breaking is induced by the VEV of another canonically normalized real scalar field 
$\chi_I$, which belongs to an appropriate representation of the intermediate gauge symmetry. For example, the field that breaks the
$\mathcal{G}_{2_L2_R4_C}$ symmetry lies in $(1,3,15)$ contained in a 210-plet of 
$SO(10)$. Similar to the previous case, the potential for $\chi_I$ is
\begin{equation}
V(\phi,\chi_I)=-\frac{1}{2}\beta_I^2 \phi^2 \chi_I^2 + \frac{a_I}{4} \chi_I^4 \ ,
\end{equation}
with the final VEV  
\begin{equation}
\langle\chi_I\rangle \equiv M_I= \frac{\beta_I}{\sqrt{a_I}} M .
\end{equation}
After incorporating the finite temperature correction $(1/2)\sigma_{\chi_I}T_H^2\chi_I^2$, the effective mass-squared of $\chi_I$ reads 
\begin{equation}
\label{effective-mass_chi1}
{m^{I}_{\mathrm{eff}}}^{\! \! \! 2}=2[\beta^2\phi^2-\sigma_{\chi_I} T_H^2] \ .
\end{equation}
Therefore, the phase transition and the formation of the associated topological defects occur for
\begin{equation} \label{form_mon_1}
 \beta_I^2\phi^2 = \left( \frac{72a_I^2}{\pi^2}+\sigma_{\chi_I}\right) T_H^2 .
\end{equation}
From this equation, we can estimate the first intermediate breaking scale $M_I$ as:
\begin{equation}\label{form_mon_2}
M_I = \sqrt{\left(\frac{72a_{I}^2}{\pi^2}+\sigma_{\chi_I} \right)} \frac{H_I}{2\pi \phi_I} \frac{M}{\sqrt{a_I}} ,
\end{equation}
where $\phi_I$ is the value of the inflaton field at the phase transition, and $H_I$ is the corresponding value of the Hubble parameter. We assume that
$a_I^2\sim 0.1$ and $\sigma_{\chi_I} \sim 1$. 

Following similar steps, we display the potential for the scalar field $\chi_{II}$ whose VEV causes the second intermediate symmetry breaking:
\begin{equation}
V(\phi,\chi_{II})=-\frac{1}{2}\beta_{II}^2 \phi^2 \chi_{II}^2 + \frac{a_{II}}{4} \chi_{II}^4 \ .
\end{equation}
The effective mass-squared for $\chi_{II}$ is
\begin{equation}
\label{effective-mass_chi2}
{m^{II}_\mathrm{eff}}^2=2[\beta^2\phi^2-\sigma_{\chi_{II}} T_H^2] \ ,
\end{equation}
and the second intermediate breaking scale is
\begin{equation}\label{breaking_scale_2}
M_{II} = \sqrt{\left(\frac{72a_{II}^2}{\pi^2}+\sigma_{\chi_{II}} \right)} \frac{H_{II}}{2\pi \phi_{II}} \frac{M}{\sqrt{a_{II}}}.
\end{equation}
Here, the phase transition occurs for $\phi=\phi_{II}$ and $H_{II}$ is the Hubble parameter at $\phi_{II}$. We also choose $a_{II}^2\sim 0.1$ and $\sigma_{\chi_{II}} \sim 1$.

At this point, it should be mentioned that the logarithmic terms in the Coleman-Weinberg potential arising from the couplings of 
$\phi$ to $\chi_I$, $\chi_{II}$ can be ignored since $\beta_I$, $\beta_{II}$ $ << \beta$. Thus, we do not include them in our analysis.

\section{Intermediate Mass Monopoles}\label{sec:monopoles}
 All GUT models predict  \cite{Lazarides:2019xai} the existence of topologically stable magnetic monopoles associated with the unification scale $M_X$. In addition, the $SO(10)$ model predicts the appearance of intermediate mass monopoles carrying two quanta of Dirac magnetic charge associated with the first intermediate breaking scale $M_I$ \cite{Lazarides:2019xai}. In the $E_6$ case, monopoles with a triple Dirac charge are generated at the first intermediate phase transition 
\cite{Lazarides:2019xai}. As we will see later, the GUT monopoles are entirely inflated away, and we will thus concentrate on the monopole production at the scale $M_I$ and compare their predicted present abundance with the results of the MACRO experiment \cite{Ambrosio:2002qq}.     
The upper bound on the monopole flux from this experiment is $2.8\times 10^{-16}$ ${\mathrm{cm}^{-2}\mathrm{s}^{-1}
\mathrm{sr}^{-1}}$. We take the lower bound (or observability threshold) on the monopole flux to be $10^{-24}$ ${\mathrm{cm}^{-2}\mathrm{s}^{-1}\mathrm{sr}^{-1}}$, below which the monopoles are too diluted to be observed. We define the monopole yield as $Y_M\equiv {n_M}/{s}$, where $n_M$ and $s$ are the  monopole number density and the entropy density respectively. The MACRO bound on the monopole flux for monopole masses $m_M\sim 10^{14}$ GeV then implies \cite{Kolb:1990vq} that the maximum allowed $Y_M$ is $Y_M^{\mathrm{max}}\sim 10^{-27}$, while the observability threshold adopted here corresponds to the minimal value of the monopole yield $Y_M^{\mathrm{min}}\sim 10^{-35}$.

 We next turn to the discussion of monopole production at $M_I$ and the subsequent evolution of their abundance \cite{Lazarides:1984pq,Lazarides:2019xai}.
We assume that the mean inter-monopole distance at production is of order ${m^{I}_{\mathrm{eff}}}^{\! \! \! -1}$, such that the monopole number density is $\simeq (1/10) {m^{I}_{\mathrm{eff}}}^{\! \! \! 3}$, where we included a numerical factor 1/10. The monopoles are subsequently diluted by the factors $\exp\left(-3N_I\right)$ and $\left(t_r/\tau\right)^2$ during inflation and inflaton oscillations respectively. Here, $t_r$ is the reheat time, which is about $0.36$ $\mathrm{GeV}^{-1}$ for 
$T_r=10^9$ GeV and for the SM spectrum, and $\tau$ is the cosmic time at the end of inflation. The monopole number density at reheating is about $({m^{I}_{\mathrm{eff}}}^{\! \! \! 3}/10) \exp(-3N_I)
(\tau/t_r)^2$, while the entropy density is $({2\pi^2}/{45})g_{*}T_r^3$. Using these estimates we can find the monopole yield $Y_M = n_M/s$ after reheating as a function of $M_I$:
 \begin{equation}
 Y_M \simeq \frac{\frac{{m^{I}_{\mathrm{eff}}}^{\! \! \! 3}}{10} \exp(-3N_I)\left(\frac{\tau}{t_r}
\right)^2}{\frac{2\pi^2}{45}g_{*}T_r^3} \ .
 \end{equation}
Here, $H_I^2 = V(\phi_I)/3 m_{\mathrm{Pl}}^2$, $N_I = (1/m_\mathrm{Pl}^2) \int_{\phi_e}^{\phi_I} V \mathrm{d}\phi/V^\prime$, and $g_{*}=106.75$ for the SM spectrum. Using the equation 
\begin{equation}\label{rollover}
3H\dot{\phi}+V'(\phi)\simeq 0 \ ,
\end{equation}
which holds during inflation to a good approximation, we can compute the time $\tau$ at the termination of inflation as follows:
\begin{equation}\label{rollover-time-1}
\tau\simeq\int_{\phi_e}^{\phi_*} \frac{3H(\phi)}{V'}) d\phi \ .
\end{equation}


\begin{table}[htbp]
\begin{center}
	\small\addtolength{\tabcolsep}{-2pt}
\begin{tabular}{| c | c | c | c | c | c | c | c | c | c |}
\hline
 \multirow{2}{*}{$\frac{V_0^{1/4}}{10^{16}\text{GeV}}$} & \multirow{2}{*}{$\frac{\tau}{10^{-12}\text{GeV}^{-1}}$} & \multirow{2}{*}{$\phi_{+}/m_{\rm Pl}$} & \multirow{2}{*}{$\phi_{-}/m_{\rm Pl}$} & ${H_+}$ & ${H_-}$ & \multirow{2}{*}{$N_+$} & \multirow{2}{*}{$N_-$} & \multirow{2}{*}{$\log_{10}\left(\frac{M_{I+}}{\text{GeV}}\right)$} & \multirow{2}{*}{$\log_{10}\left(\frac{M_{I-}}{\text{GeV}}\right)$} \\
 \cline{5-6}
 &&&&\multicolumn{2}{c|}{({\tiny{$ 10^{13}$ GeV}})} & &&& \\
 \hline
  1.51 & 1.38 & 14.41 & 13.07 & 3.40 & 3.91 & 9.8 & 16.2 & 13.30 & 13.40 \\
  1.59 & 1.32 & 16.04 & 14.67 & 3.54 & 4.10 & 9.9 & 16.2 & 13.30 & 13.41 \\
  1.66 & 1.26 & 17.91 & 16.51 & 3.67 & 4.28 & 9.9 & 16.2 & 13.31 & 13.41 \\
  1.74 & 1.22 & 20.05 & 18.62 & 3.78 & 4.45 & 9.9 & 16.2 & 13.31 & 13.41 \\
  1.82 & 1.18 & 22.51 & 21.04 & 3.88 & 4.59 & 9.9 & 16.2 & 13.31 & 13.41 \\
 \hline
\end{tabular}
\caption{Values of the various parameters (indicated by a subscript $+$) corresponding to the MACRO bound on the flux of
monopoles formed at the scale $M_I$ and their values (indicated by a subscript $-$) corresponding to the adopted observability threshold for the monopole flux.}
\label{Infl_bounds_1}
\end{center}
\end{table}

In Table~\ref{Infl_bounds_1}, we present the minimal required numbers of $e$-foldings $N_{+}$ which must follow the monopole production so that the MACRO bound on the monopole flux is satisfied. We also show the corresponding lower bounds $M_{I+}$ on $M_I$, as well as the corresponding values of the inflaton $\phi_{+}$, and the Hubble parameter $H_{+}$ at monopole production. 
We also estimate the values of these parameters (indicated by a subscript $-$) corresponding to the threshold for observability.
 
 
\begin{figure}[htbp]
\subfloat[$SO(10)$.]
{
\includegraphics[scale=0.355]{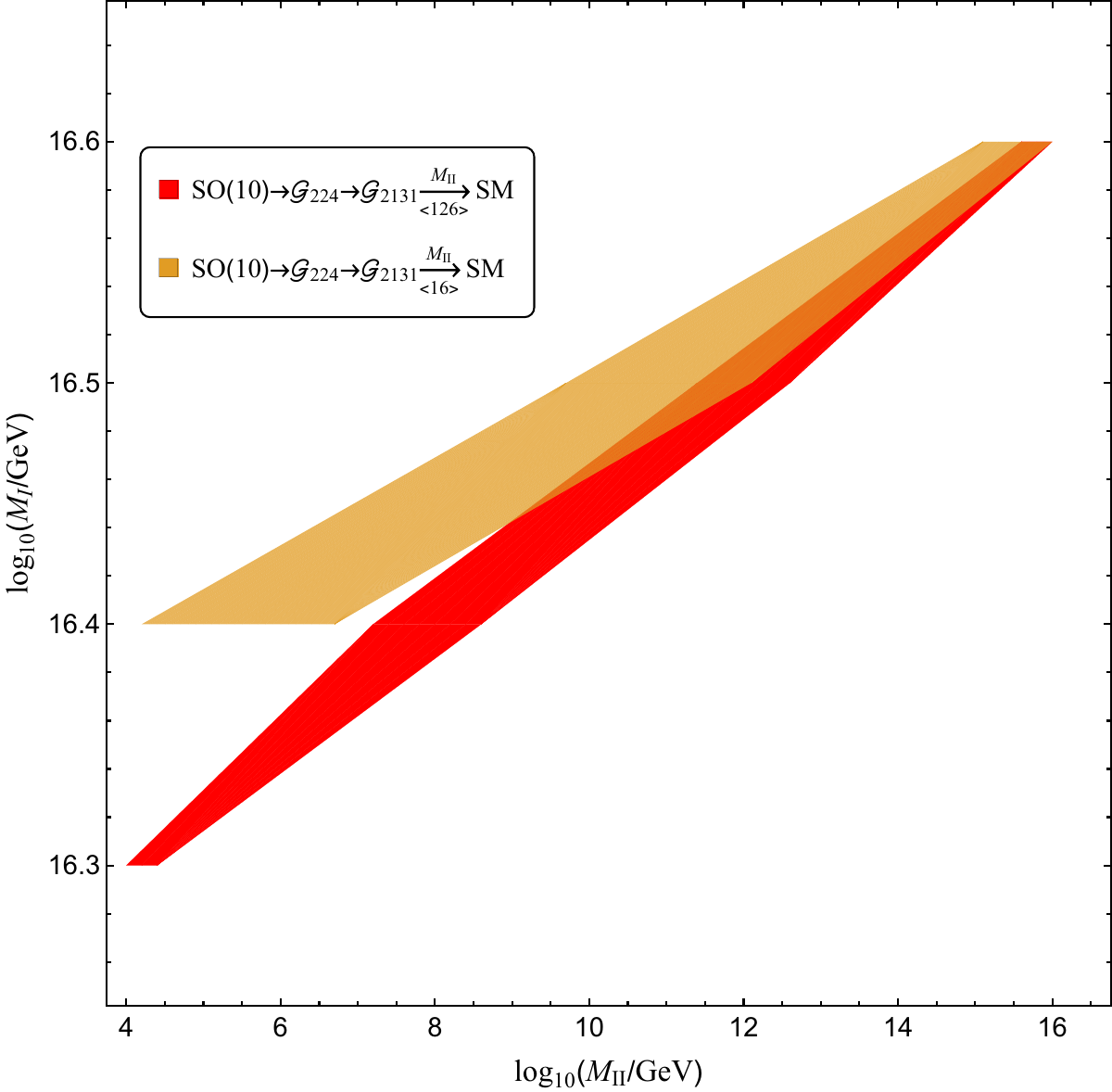}
}
\hspace{8mm}
\subfloat[$E_6$.]
{
\includegraphics[scale=0.35]{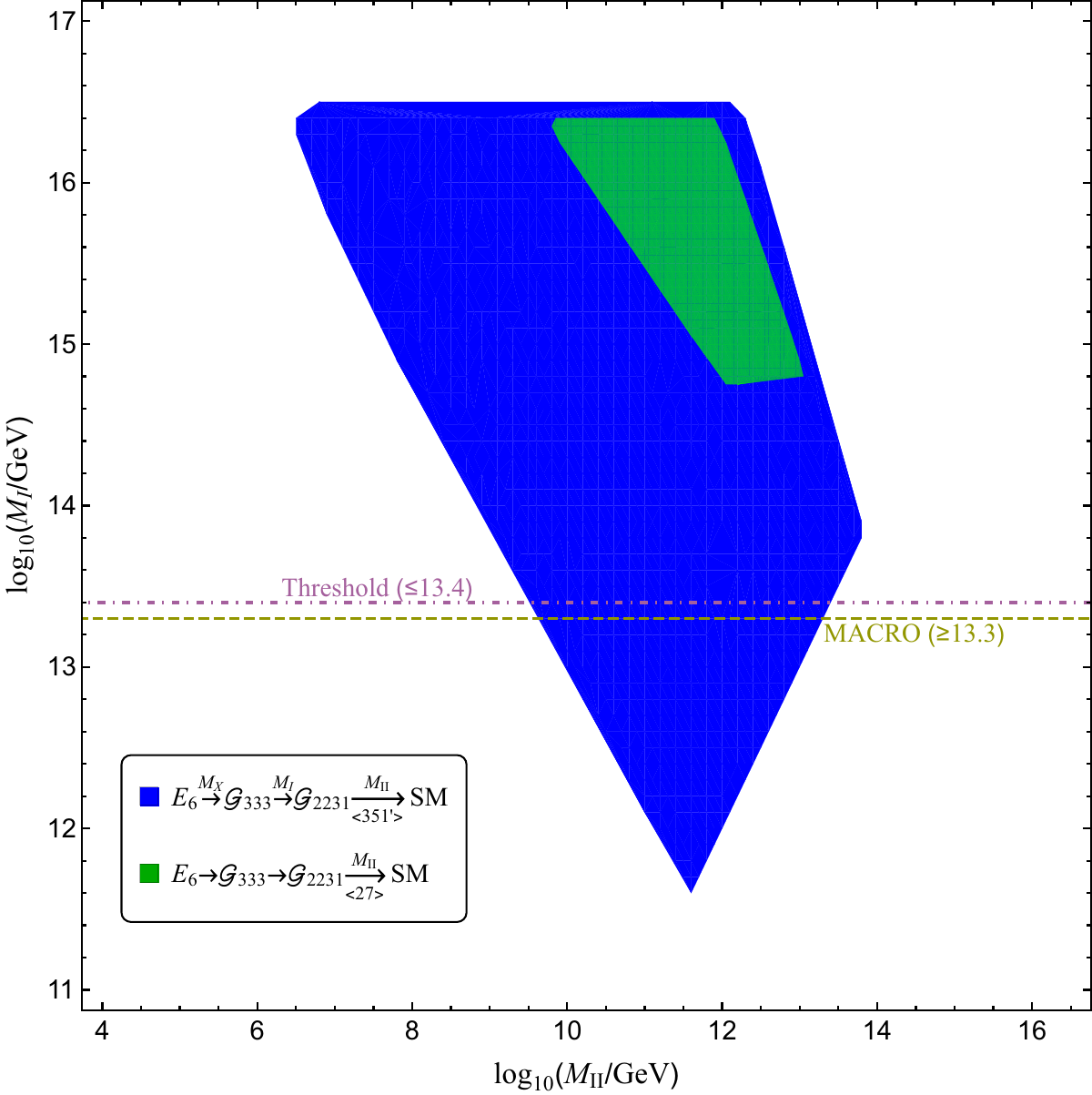}
}

\caption{Intermediate breaking scales $M_I$ and $M_{II}$ for $\log_{10}(M_X/\mathrm{GeV}) \in [16.55, 16.63]$ in the case of
$SO(10)$ and $\log_{10}(M_X/\mathrm{GeV}) \in [16.43, 16.51]$ in the case of $E_6$ with successful inflation based on a Coleman-Weinberg potential. We also show the two bounds $M_{I+}$ and $M_{I-}$ on $M_I$ derived by using Eq.~(\ref{form_mon_2}). Namely, the horizontal lines at $\log_{10}(M_{I+}/\mathrm{GeV}) = 13.3$ (dashed) and $\log_{10}(M_{I-}/\mathrm{GeV}) = 13.4$ (dot-dashed) represent the MACRO bound and the upper bound on $M_I$ for observability of the monopole flux (for $M_I$'s above this value, the monopoles are too diluted to be observed).}\label{cw_range}
\end{figure}


In Fig.~\ref{cw_range}, we show the allowed ranges of the intermediate scales $M_I,\; M_{II}$ which are consistent with successful inflation based on a Coleman-Weinberg potential for the four GUT scenarios considered with the unification scale $M_X$ restricted in the range $\log_{10}(M_X/\mathrm{GeV}) \in [16.55, 16.63]$ for $SO(10)$, and $\log_{10}(M_X/\mathrm{GeV}) \in [16.43, 16.51]$ for 
$E_6$. It should be mentioned that the unification scale is perfectly consistent with the proton lifetime bounds suggested by the present Super-Kamiokande results as well as the expected sensitivity of the future Hyper-Kamiokande experiment. We note that out of the four scenarios only the $E_6$ unified model with $\mathbb{Z}_2$ strings can yield an observable flux of triply charged monopoles produced at the intermediate scale $M_I$. The MACRO bound excludes a considerable part of the available (blue) region for this model -- see Fig.~\ref{cw_range}.  It also suggests that the monopoles corresponding to the unification scale $M_X$ are inflated away in all cases.

\section{Intermediate Scale Strings and Gravity Waves}\label{sec:strings}

The $\mathbb{Z}_2$ cosmic strings are formed \cite{Lazarides:2019xai} at $M_{II}$ when the parent symmetry is broken through the VEV of a sub-multiplet of $351'$ or $\overline{126}$ in $E_6$ or $SO(10)$ respectively.
The mean inter-string distance $d_s$, i.e. the scale of the network, at formation is expected to be 
\begin{equation}\label{eq:ds}
d_s\simeq p\; {m^{II}_{\mathrm{eff}}}^{-1},
\end{equation}
where $p\simeq 2$ is a geometric factor. For generic values of $M_{II}$, these strings will be formed during inflation. But, for suitably lower $M_{II}$ values, the strings can appear after the end of inflation, either during the inflaton oscillations or even after reheating. 
 
We first investigate the situation with $M_{II}$ large enough so that the string formation takes place during the inflationary era. From Eq.~(\ref{breaking_scale_2}), we can find the lower bound on $M_{II}$ for this to happen: 
\begin{equation}
\label{M_II_end_infl}
M_{II} > \sqrt{\left(\frac{72\alpha_{II}^2}{\pi^2}+\sigma_{\chi_{II}} \right)} \frac{H(\phi_e)}{2\pi} \frac{M}{\phi_e\sqrt{\alpha_{II}}} \ ,
\end{equation}
with $H(\phi_e)=\sqrt{V(\phi_e)/3m_{\rm Pl}^2}$. The inflaton value $\phi=\phi_{II}$ at the phase transition can be computed again from Eq.~(\ref{effective-mass_chi2}). 
We then determine the effective scalar mass using Eq.~(\ref{effective-mass_chi2}) and the mean inter-string distance from Eq.~(\ref{eq:ds}). During inflation, $d_s$ is scaled by a factor $\exp(N_{II})$, where $N_{II}=(1/m_\mathrm{Pl}^2)\int_{\phi_e}^{\phi_{II}} V \mathrm{d}\phi/V^\prime$ is the number of $e$-foldings after the string formation.
The inter-string distance gets further scaled by two additional  factors, namely by $(t_r/\tau)^{2/3}$ during the period of inflaton oscillations and by $T_r/T_0$ from reheating to the present time, where $T_0=2.35 \times 10^{-13}$ GeV is the present cosmic microwave background temperature.
Including these factors, we estimate the present value of $d_s$
\begin{equation}\label{hrznentryscale}
d_s\simeq p~ {m^{II}_\mathrm{eff}}^{-1}(\phi_{II})\exp(N_{II})\left(\frac{t_r}{\tau}\right)^{\frac{2}{3}} 
\frac{T_r}{T_0} \ .
\end{equation}
For strings to enter the present horizon, i.e. not to be inflated away, the inter-string distance in Eq.~(\ref{hrznentryscale}) should be smaller than the present horizon size $3t_0$, where $t_0=6.62 \times 10^{41} \ \mathrm{GeV}^{-1}$ is the present cosmic time.
  
The dimensionless string tension  $G\mu$ is given by 
\begin{equation}\label{Gmu}
G\mu\simeq\frac{1}{8}\left(\frac{M_{II}}{m_{\rm Pl}}\right)^2 \ ,
\end{equation}
where $G$ and $\mu$ are Newton's constant and  the string tension, i.e. the string mass per unit length, respectively. Here, our assumption is  that these strings are close to the Bogomol'nyi limit of the Abelian Higgs model
\cite{Bevis:2006mj,Bevis:2007qz,Bevis:2007gh}. From PTA \cite{Shannon:2015ect}, we know  that $G\mu \lesssim 1.5 \times 10^{-11}$ \cite{Blanco-Pillado:2017rnf}, which implies that $M_{II}\lesssim 2.7 \times 10^{13}$ GeV. (For recent developments see 
Refs.~\cite{Arzoumanian:2020vkk,Ellis:2020ena,Buchmuller:2020lbh,Pol:2020igl}.) As we will see later, this bound is strictly applicable only to those strings that enter the horizon before $t\simeq 10 
\Gamma G\mu t_{eq}$ ($\Gamma$ is a numerical factor of order 50), and certainly not for strings entering the horizon after the equidensity time $t_{eq}\simeq 2.253\times 10^{36}~{\rm GeV}^{-1}$, where the energy densities of radiation and matter coincide. 

The mean inter-string distance at a cosmic temperature $T$ after reheating and before the equidensity point can be estimated from Eq.~(\ref{hrznentryscale}) with $T_0$ replaced by $T$, where
\begin{equation}\label{time-rad-dom}
T^2=\sqrt{\frac{45}{2\pi^2}}g_*^{-1/2}\frac{m_{\rm Pl}}{t} \ ,
\end{equation} 
with the appropriate value of $g_*$ for the relevant temperature range. Equating this inter-string distance with the horizon distance $2t$, we can calculate the scale $M_{II}$ for which the strings enter the horizon at any given cosmic time $t$ during radiation dominance. After horizon entrance the long strings chop each other and inter-commute generating loops of typical size 
$\ell\simeq t/10$ at any subsequent time $t$ \cite{Blanco-Pillado:2013qja,Blanco-Pillado:2017oxo}. These loops eventually decay \cite{Vilenkin:2000jqa} into gravity waves at $t\simeq \ell/\Gamma G\mu$, providing the major contribution to the stochastic background.
Strings that enter the horizon before $10 \Gamma G\mu t_{eq}$ give rise to a complete spectrum of loops generated between this time and $t_{eq}$ and decaying after $t_{eq}$. These loops which are created during radiation dominance and decay during matter dominance generate \cite{Sousa:2020sxs} the overall peak of the stochastic gravity waves which lies at low frequencies and is restricted by the PTA bound.

 In order to compute the scale $M_{II}$ corresponding to strings entering the horizon after $t_{eq}$, we need to solve the inequality
\begin{equation}
3t_{eq} < p~ m_{eff}^{-1}\exp(N_{II})\left(\frac{t_r}{\tau}\right)^{\frac{2}{3}} (\frac{T_r}{T_{eq}}) \ ,
\end{equation}
where $T_{eq}=9.45\times 10^{-10}$ GeV is the temperature at $t_{eq}$. These strings may generate
\cite{Sousa:2020sxs}
an insignificant low frequency peak in the gravity wave spectrum which is overshadowed by the overall peak and thus they are not important for the PTA bound. 

\begin{figure}[htbp]
\begin{center}
\includegraphics[scale=0.6]{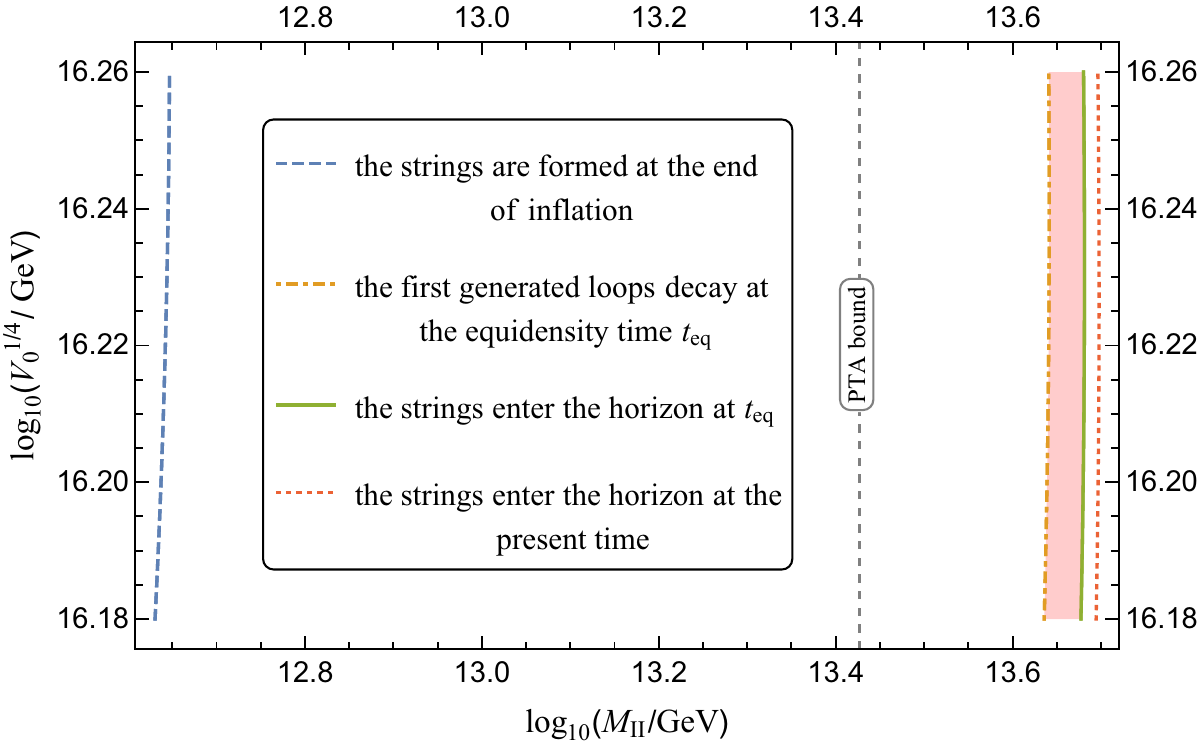}
\end{center}
\caption{Intermediate breaking scales $M_{II}$ for the unification scale $\log_{10}(V_0^{1/4}/\mathrm{GeV}) \in[16.18, 16.26]$ for successful inflation with Coleman-Weinberg potential for different cases: 
1. strings are formed at the end of inflation, 2. earliest loops decay at the equidensity time 
$t_{eq}$, 3. strings enter the horizon at $t_{eq}$, and 4. strings enter the horizon at the present time. The dashed black line corresponding to the PTA bound is also 
shown.}\label{Strngplt}
\end{figure}

In Fig.~\ref{Strngplt}, we show the values of the second intermediate scale $M_{II}$ which correspond to strings generated during inflation that re-entered the horizon during different eras of the universe, consistent with successful inflation (see Table~\ref{tab:Infl_para}). 

Let us summarize the various regions in Fig.~\ref{Strngplt}:

\begin{itemize}
\item For $\log_{10}(M_{II}/\mathrm{GeV})\gtrsim 12.64$, the phase transition takes place and strings are generated before the end of inflation.
\item For $\log_{10}(M_{II}/\mathrm{GeV}) \lesssim 13.64$, the strings enter the horizon before
$t\simeq 10 \Gamma G\mu t_{eq}$, and thus the loops generated after this time and before 
$t_{eq}$ are present. These loops generate \cite{Sousa:2020sxs} a significant low frequency peak in the spectrum of stochastic gravity waves and the restriction from the PTA bound is expected to hold in this case. Loops created before $t\simeq 10 \Gamma G\mu t_{eq}$ decay during radiation dominance and contribute \cite{Sousa:2020sxs} to the plateau of the spectrum.
\item
For $13.64 \lesssim \log_{10}(M_{II}/\mathrm{GeV}) \lesssim 13.68$, the strings enter the horizon after 
$t\simeq 10 \Gamma G\mu t_{eq}$ and before $t_{eq}$. Only part of the loops that are generated during radiation dominance and decay after $t_{eq}$ are present. Consequently, the low frequency peak in the spectrum gradually fades away as $M_{II}$ increases in this region. This region has been shaded in Fig.~\ref{Strngplt}. Only the part corresponding to lower $M_{II}$ values may be excluded by the PTA bound.
\item For $13.68 \lesssim \log_{10}(M_{II}/\mathrm{GeV}) \lesssim 13.70$, the strings enter the horizon after $t_{eq}$ and the significant low frequency peak in the gravity wave spectrum is absent. Consequently, there is no restriction from the PTA experiment. 
\item For $\log_{10}(M_{II}/\mathrm{GeV})\gtrsim 13.70$, the strings never enter the horizon and thus again no restriction arises.
\end{itemize}

\section{Phase Transition after Inflation}
\label{sec:afterinf}
The phase transition occurs during inflaton oscillations or even after reheating if $M_{II}$ does not satisfy the inequality in Eq.~(\ref{M_II_end_infl}), i.e. if $M_{II}\lesssim 4.36 \times 10^{12}$ GeV. The PTA bound is certainly well satisfied in this case. Strings produced after the end of inflation always remain inside the post-inflationary horizon. It is interesting to note that the causality criterion forbids the inter-string distance to be bigger than the horizon size. Long strings reach the scaling solution quickly with one string segment per horizon and start generating loops almost instantaneously.
 
At the beginning of inflaton oscillations, the corrections to the mass-squared of the field $\chi_{II}$ are not dominated by the temperature corrections from the ``new'' radiation, but by the Hubble parameter $H_\phi$ from the energy density $\rho_\phi(t)$ of the oscillating inflaton 
\cite{Dine:1983ys}. Assuming that these oscillations are quadratic, we have (for a review see Ref.~\cite{Lazarides:2001zd})
\begin{align}
\rho_\phi(t) = \rho_e \left(\frac{t}{\tau}\right)^{-2}\exp[-\Gamma_\phi (t-\tau)],
\label{eq:rhophit}
\end{align}
where $\Gamma_\phi\simeq 2.8$ GeV is the inflaton decay width.
Therefore, right after the end of inflation, the correction to the $\chi_{II}$ mass-squared term is  
\begin{equation}
\frac{1}{2}\sigma \left(\frac{H_\phi}{2\pi}\right)^2 \chi_{II}^2,
\label{afterinf} 
\end{equation}
where $H_\phi=\sqrt{\rho_{\phi}/3m_\mathrm{Pl}^2}$ and the corresponding Hawking temperature is $T_H=H_\phi/2\pi$. Here, we set $\sigma=1$ so that continuity of the correction between the inflationary and the oscillatory era is guaranteed. Using the correction in Eq.~(\ref{afterinf}) and following the analysis of Sec.~\ref{sec:phasetrans}, one can then calculate the value of $\chi_{II}$ at the minima of the potential with the mean value of $\phi^2$  $\simeq M^2$, since $\phi$ oscillates about $M$ with an amplitude smaller than $M$. The potential difference $\Delta V$ between the local maximum at $\chi_{II}= 0$ and these minima, as well as the effective mass $m_{\mathrm{eff}}$ of $\chi_{II}$ at the minima are also estimated. The Ginzburg criterion for this case then takes the form: 
\begin{equation}
\frac{4\pi}{3}m_{\mathrm{eff}}^{-3}\Delta V > T_H \Rightarrow \beta_{II}^2 M^2 > \left( \frac{72a_{II}^2}{\pi^2}+\sigma\right) T_H^2.
\label{ginzburg2}
\end{equation}
The value of $T_H$ (and thus $\rho_\phi$) at which the phase transition takes place for given $M_{II}$ can be calculated by saturating this inequality.

The new radiation energy density $\rho_r$ is given by (for a review see Ref.~\cite{Lazarides:2001zd}) 
\begin{align}\label{temp_rad_rh_2}
\rho_r(t)=\rho_e \left(\frac{t}{\tau}\right)^{-8/3}\int_\tau^t \left(\frac{t'}{\tau}\right)^{2/3}\exp[-\Gamma_\phi (t'-\tau)]\mathrm{d}t' \ ,
\end{align}
and its temperature $T$ can be found from $\rho_r=({\pi^2}/{30})g_*T^4$. As it turns out, $T$ soon becomes larger than $T_H$ and dominates the correction to the mass-squared term of $\chi_{II}$, which takes the form $(1/2)\sigma^\prime T^2 \chi_{II}^2$. Here $\sigma^\prime$ is, in principle, different from $\sigma$, but for simplicity we take it again equal to unity. Needless to say that, in $m_\mathrm{eff}^{2}$ and $\Delta V$, 
$T_H$ and $\sigma$ should be replaced by $T$ and $\sigma^\prime$ respectively, and $\phi^2$ by $M^2$. The Ginzburg criterion is then as in Eq.~(\ref{ginzburg2}) with $T_H$ replaced by $T$ and $\sigma$ replaced by $\sigma^\prime$. The temperature $T$ of the new radiation at which the transition takes place for given $M_{II}$ is again calculated by saturating the Ginzburg criterion. It is important to note that there is continuity of the $\chi_{II}$ mass-squared correction between the regimes where this correction is dominated by $T_H$ or $T$. The latter regime smoothly extends even to the period after reheating. 
\begin{table}[htbp]
	\begin{center}
		\small\addtolength{\tabcolsep}{-1pt}
		\begin{tabular}{| c | c | c | c | c | c | c |}
			\hline
			\multirow{2}{*}{\parbox{1.2 cm}{$\frac{V_0^{1/4}}{10^{16}\text{GeV}}$\\ }} & \multirow{2}{*}{\parbox{2.3 cm}{$H_\phi (10^{12} \ \mathrm{GeV})$ \\ at $T_H = T$}} & \multirow{2}{*}{\thead{$T$ $(10^{12}~{\rm GeV})$\\ at $T_H=T$}} & \multicolumn{2}{c|}{$\log_{10}(M_{II}/\mathrm{GeV})$ at} & \multicolumn{2}{c|}{$t \ (10^{-12} \ \mathrm{GeV}^{-1})$ at} \\
			\cline{4-7}
			& & & end of inflation & $T_H = T$ & end of inflation & $T_H = T$ \\
			\hline
			1.51 & 8.92 & 1.42 & 12.63 & 12.57 & 1.38 & 1.49 \\
			1.59 & 9.06 & 1.44 & 12.64 & 12.58 & 1.32 & 1.42 \\
			1.66 & 9.16 & 1.46 & 12.64 & 12.58 & 1.26 & 1.37 \\
			1.74 & 9.23 & 1.47 & 12.65 & 12.58 & 1.22 & 1.33 \\
			1.82 & 9.27 & 1.48 & 12.65 & 12.59 & 1.18 & 1.29 \\
			\hline
		\end{tabular}
		\caption{Hubble parameter $H_\phi$ from the oscillating inflaton and the temperature of the new radiation with $T_H = T$ for various $V_0^{1/4}$ values corresponding to successful inflation. We also show the breaking scale $M_{II}$ and the cosmic time $t$ at the end of inflation and at $T_H = T$ for comparison. After the cosmic time at $T_H = T$, the new radiation dominates over the Hawking temperature $T_H$ from field oscillations.}
		\label{tab:TH_T}
	\end{center}
\end{table}

\begin{figure}[htbp]
	\begin{center}
		\includegraphics[scale=0.6]{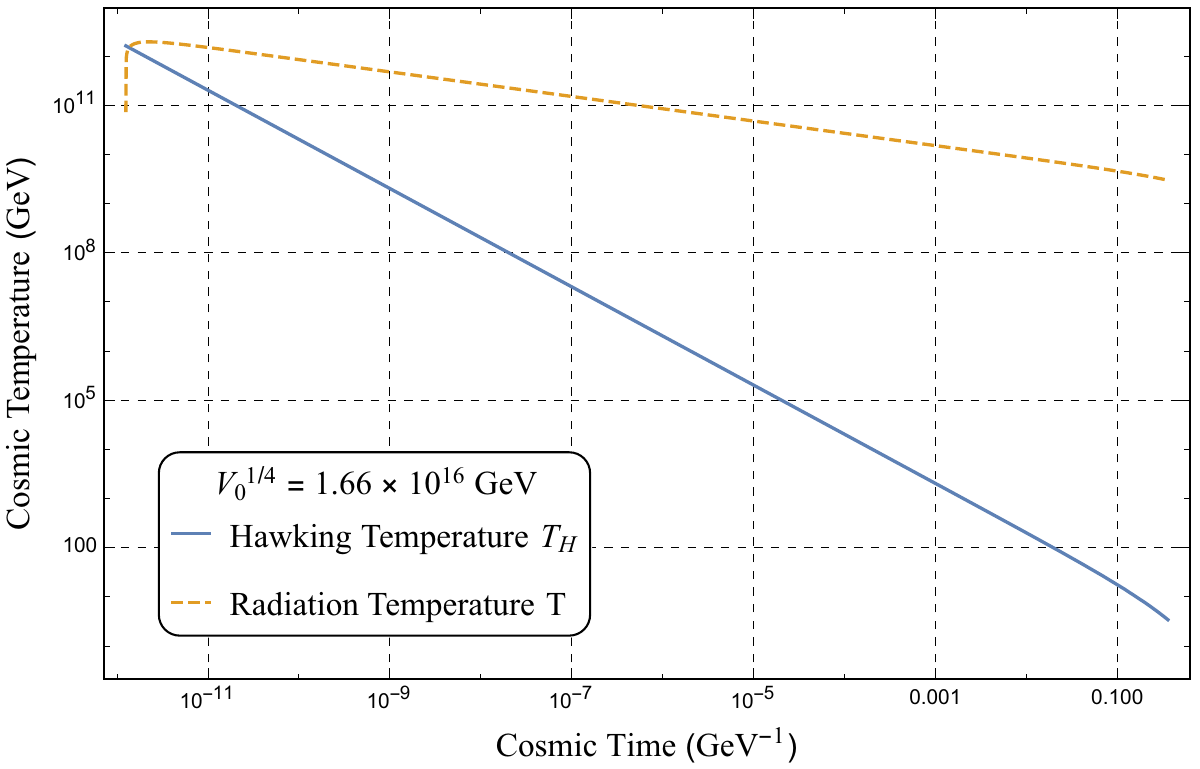}
	\end{center}
	\caption{Hawking Temperature $T_H=H_\phi/2\pi$, with $H_\phi$ being the Hubble parameter from the energy density $\rho_\phi$ of the inflaton oscillations, and new radiation temperature $T$ versus cosmic time for successful inflation with 
$V_0^{1/4}=1.66\times 10^{16}$ GeV. The new radiation temperature $T$ starts dominating over $T_H$ after cosmic time $1.37\times 10^{-12} \ \mathrm{GeV}^{-1}$, and inflation ends at 
$\tau = 1.26\times 10^{-12} \ \mathrm{GeV}^{-1}$.}\label{fig:TH_T}
\end{figure}

In order to find the limiting value of $M_{II}$ which separates the two regimes where the correction to the $\chi_{II}$ mass-squared is dominated by the oscillating inflaton or the new radiation, we compare $T_H$ and $T$ calculated by using Eq.~(\ref{eq:rhophit}) and Eq.~(\ref{temp_rad_rh_2}) respectively. We then find the cosmic time at which these two temperatures coincide and their common value. Saturating the Ginzburg criterion in Eq.~(\ref{ginzburg2}), we finally estimate the limiting value of $M_{II}$. The value of the Hubble parameter $H_\phi$ from inflaton oscillations and the cosmic temperature for $T_H = T$ are given in Table \ref{tab:TH_T} for successful inflation. We also provide in this table the breaking scale $M_{II}$ and the cosmic time $t$ at the end of inflation and at $T_H = T$ for comparison. After the cosmic time at which $T_H = T$, the temperature of the new radiation starts dominating over the Hawking temperature $T_H$. The variation of $T_H$ and $T$ with the cosmic time $t$ is shown in Fig.~\ref{fig:TH_T} for $V_0^{1/4}=1.66\times 10^{16}$ GeV.

Loops that are produced during any phase transition occurring within the era of inflaton oscillations decay much earlier than the equidensity time $t_{eq}$. Indeed, a phase transition taking place at the reheat time $t_r\simeq 0.36~{\rm GeV}^{-1}$ corresponds to a breaking scale around $M_{II}\simeq 2.34\times 10^9$ GeV. The lifetime of the loops of size $\sim t_r/10$ generated at this transition is 
$6.43\times 10^{15}$ $\mathrm{GeV}^{-1}$ and thus these loops contribute to the plateau in the gravity wave spectrum. During radiation dominance, a loop produced at cosmic temperature $T$ and time $t$ decays at $t_{eq}$ if $t\simeq 10(\Gamma G\mu)t_{eq}$, where $t$ and $G\mu$ are estimated from Eqs.~(\ref{time-rad-dom}) and (\ref{Gmu}), respectively. The minimum value of the symmetry breaking scale $M_{II}$ for 
which the loops generated during the corresponding phase transition in a radiation dominated universe contribute to the sharp peak in the gravity wave spectrum can then be found from the Ginzburg criterion and turns out to be $2.6\times 10^4$ GeV.

\section{Conclusions}
\label{sec:conclusion}
We have explored in this paper the appearance and subsequent evolution of topologically stable magnetic monopoles and cosmic strings in realistic non-supersymmetric SO(10) and $E_6$ GUTs. As an important first step we perform a comprehensive study of GUT symmetry breaking with two intermediate scales that is compatible with gauge coupling unification and proton decay limits. In turn, this allows us to identify the monopoles and strings associated with the GUT and intermediate scale symmetry breakings. Topological defects with intermediate scales are of special interest and, to keep things realistic, we explore their evolution within the context of an inflationary universe. We highlight  models which predict the presence of an observable number density of primordial monopoles with mass $\sim 10^{13}-10^{14}$ GeV and cosmic strings with the string tension parameter $G \mu \sim 10^{-11}-10^{-10}$ that have survived an inflationary epoch. The impact of inflation on the stochastic gravitational background radiation emitted by strings is also discussed. Finally, we note that $G\mu$ values lying in a wide range $\sim 10^{-10} - 10^{-20}$ will be probed by a variety of proposed experiments including LISA \cite{Bartolo:2016ami, amaroseoane2017laser}, SKA \cite{5136190, Janssen:2014dka}, BBO \cite{Crowder:2005nr, Corbin:2005ny}, and ET \cite{Mentasti:2020yyd}.

\section{Acknowledgment}
The work of J.C. and R.M. is supported by the Science and Engineering Research Board, Government of India, under the agreements SERB/PHY/2016348 (Early Career Research Award) and SERB/PHY/2019501
(MATRICS).  The work of G.L. and Q.S. is supported by the Hellenic Foundation for Research
and Innovation (H.F.R.I.) under the ``First Call for H.F.R.I. Research Projects to support Faculty Members and Researchers and the procurement of high-cost research equipment grant'' (Project 
Number:2251).  Q.S. thanks Nefer Vedat \c{S}eno\u{g}uz  for a useful discussion.

\appendix
\section*{Appendix: RGEs for the Two Breaking Chains of $SO(10)$}\label{Appendix}
\addcontentsline{toc}{section}{Appendix: RGEs for the Two Breaking Chains of $SO(10)$}

\subsection*{A.1\quad The RGEs and $\beta$-coefficients for the Breaking Chain in 
Sec.~\ref{SO10_A}}

	               From $M_{II}$ to $M_I$:\\		
\begin{align*}
\mu \frac{\mathrm{d}g_{2L}}{\mathrm{d}\mu} = &\frac{1}{(4\pi)^2}\left( - 3 g_{2L}^{3}\right)
+\frac{1}{(4\pi)^4}\left(8 g_{2L}^{5} + 12 g_{2L}^{3} g_{3C}^{2} + \frac{3 g_{2L}^{3}}{2} g_{XX}^{2} + g_{2L}^{3} g_{RR}^{2}\right. \\& \left. + g_{2L}^{3} g_{RX}^{2} + \frac{3 g_{2L}^{3}}{2} g_{XR}^{2}\right) \ ,
\end{align*}

\begin{align*}
\mu \frac{\mathrm{d}g_{3C}}{\mathrm{d}\mu} = &\frac{1}{(4\pi)^2}\left(- 7 g_{3C}^{3}\right)+ \frac{1}{(4\pi)^4}\left(\frac{9 g_{2L}^{2}}{2} g_{3C}^{3} - 26 g_{3C}^{5} + \frac{g_{3C}^{3}}{2} g_{XX}^{2} + \frac{3 g_{3C}^{3}}{2} g_{RR}^{2} \right. \\& \left.  + \frac{3 g_{3C}^{3}}{2} g_{RX}^{2} + \frac{g_{3C}^{3}}{2} g_{XR}^{2}\right) \ ,
\end{align*}

\begin{align*}
\mu \frac{\mathrm{d}g_{RR}}{\mathrm{d}\mu} = & \frac{1}{(4\pi)^2}\left( \frac{53}{12} g_{RR}^{3} - \frac{\sqrt{6}}{12}g_{XR} g_{RR}^{2} + \frac{53}{12} g_{RR} g_{RX}^{2} + \frac{33}{8} g_{RR} g_{XR}^{2} \right. \\
& \left. - \frac{g_{XX}}{24} g_{RR} \sqrt{6} g_{RX} + \frac{33}{8} g_{XX} g_{RX} g_{XR} - \frac{g_{XR}}{24} \sqrt{6} g_{RX}^{2}\right)\\
& + \frac{1}{(4\pi)^4}\left(3 g_{2L}^{2} g_{RR}^{3} + 12 g_{3C}^{2} g_{RR}^{3} + \frac{15}{8} g_{XX}^{2} g_{RR}^{3} + \frac{17}{4} g_{RR}^{5}  \right. \\
& + \frac{17}{2} g_{RR}^{3} g_{RX}^{2} + \frac{17}{4} g_{RR} g_{RX}^{4} +3 g_{2L}^{2} g_{RR} g_{RX}^{2} + 12 g_{3C}^{2} g_{RR} g_{RX}^{2} \\
&  + 4 g_{3C}^{2} g_{RR} g_{XR}^{2} + \frac{45}{4} g_{RR}^{3} g_{XR}^{2} + \frac{65}{16} g_{RR} g_{XR}^{4} - \frac{ \sqrt{6}}{2}g_{XR} g_{RR}^{4} \\
&  - \frac{3 \sqrt{6}}{4} g_{RR}^{2} g_{XR}^{3} - \frac{ \sqrt{6}}{8}g_{XR} g_{RX}^{4} - \frac{3 \sqrt{6}}{16} g_{RX}^{2} g_{XR}^{3} + \frac{9}{2} g_{RR} g_{2L}^{2} g_{XR}^{2} \\
& + \frac{45}{8} g_{RR} g_{XX}^{2} g_{RX}^{2} + \frac{45}{8} g_{XX} g_{RX}^{3} g_{XR} + \frac{15}{2} g_{RR} g_{RX}^{2} g_{XR}^{2}\\
&+4 g_{3C}^{2} g_{XX} g_{RX} g_{XR} + \frac{65}{16} g_{XX}^{3} g_{RX} g_{XR} + \frac{65}{16} g_{RR} g_{XX}^{2} g_{XR}^{2} \\
& + \frac{65}{16} g_{XX} g_{RX} g_{XR}^{3}- \frac{9 \sqrt{6}}{16} g_{XX}^{2} g_{RX}^{2} g_{XR} - \frac{3 \sqrt{6} }{8} g_{XX}g_{RR} g_{RX}^{3}\\
& - \frac{5 \sqrt{6}}{8} g_{RR}^{2} g_{RX}^{2} g_{XR} + \frac{9}{2} g_{XX} g_{2L}^{2} g_{RX} g_{XR} - \frac{3  \sqrt{6}}{16}g_{RR} g_{XX}^{3} g_{RX} \\
& - \frac{3 \sqrt{6}}{8} g_{XX}^{2} g_{RR}^{2} g_{XR} - \frac{3 \sqrt{6} }{8} g_{XX}g_{RR}^{3} g_{RX}\\
& \left. +\frac{105}{8} g_{XX} g_{RR}^{2} g_{RX} g_{XR} - \frac{15 \sqrt{6}}{16} g_{XX}  g_{RR} g_{RX} g_{XR}^{2}\right) \ ,
\end{align*}

\begin{align*}
\mu \frac{\mathrm{d}g_{RX}}{\mathrm{d}\mu} = & \frac{1}{(4\pi)^2}\left(\frac{33}{8} g_{XX}^{2} g_{RX} - \frac{\sqrt{6}}{12} g_{XX} g_{RX}^{2} + \frac{53}{12} g_{RR}^{2} g_{RX} + \frac{53}{12} g_{RX}^{3} \right.\\
& \left.- \frac{\sqrt{6}}{24} g_{XX} g_{RR}^{2} + \frac{33}{8} g_{XX} g_{RR} g_{XR} - \frac{\sqrt{6}}{24} g_{RR} g_{RX} g_{XR}\right) \\
&+ \frac{1}{(4\pi)^4}\left( +3 g_{2L}^{2} g_{RX}^{3} + 12 g_{3C}^{2} g_{RX}^{3} + \frac{17 g_{RR}^{4}}{4} g_{RX} + \frac{17}{2} g_{RR}^{2} g_{RX}^{3}\right.\\
& + \frac{17}{4} g_{RX}^{5} + \frac{15}{8} g_{RX}^{3} g_{XR}^{2} + 3 g_{2L}^{2} g_{RR}^{2} g_{RX} + 4 g_{3C}^{2} g_{XX}^{2} g_{RX} \\
&  + 12 g_{3C}^{2} g_{RR}^{2} g_{RX} + \frac{65 g_{XX}^{4}}{16} g_{RX} + \frac{45}{4} g_{XX}^{2} g_{RX}^{3} - \frac{3 \sqrt{6}}{16} g_{XX}^{3} g_{RR}^{2} \\
& - \frac{3 \sqrt{6}}{4} g_{XX}^{3} g_{RX}^{2} - \frac{\sqrt{6}}{8} g_{XX} g_{RR}^{4} - \frac{\sqrt{6}}{2} g_{XX} g_{RX}^{4} + \frac{9}{2} g_{2L}^{2} g_{XX}^{2} g_{RX}\\
& + \frac{15}{2} g_{XX}^{2} g_{RR}^{2} g_{RX} + \frac{45}{8} g_{XX} g_{RR}^{3} g_{XR} + \frac{45}{8} g_{RR}^{2} g_{RX} g_{XR}^{2}\\
&+4 g_{3C}^{2} g_{XX} g_{RR} g_{XR} + \frac{65}{16} g_{RR} g_{XX}^{3} g_{XR} + \frac{65 }{16}g_{XX}^{2} g_{RX} g_{XR}^{2} + \frac{65}{16} g_{XX} g_{RR} g_{XR}^{3}\\
&- \frac{5 \sqrt{6} }{8} g_{XX}g_{RR}^{2} g_{RX}^{2} - \frac{9}{16} g_{XX} \sqrt{6} g_{RR}^{2} g_{XR}^{2} - \frac{3 \sqrt{6} }{8} g_{XX}g_{RX}^{2} g_{XR}^{2}\\
&+\frac{9}{2} g_{XX} g_{2L}^{2} g_{RR} g_{XR} - \frac{3 \sqrt{6}}{8} g_{RR}^{3} g_{RX} g_{XR} - \frac{3  \sqrt{6}}{8}g_{RR} g_{RX}^{3} g_{XR} \\
&\left. - \frac{3  \sqrt{6}}{16}g_{RR} g_{RX} g_{XR}^{3} - \frac{15\sqrt{6}}{16} g_{RR} g_{XX}^{2} g_{RX} g_{XR} + \frac{105}{8} g_{XX} g_{RR} g_{RX}^{2} g_{XR}\right) \ ,
\end{align*}

\begin{align*}
\mu \frac{\mathrm{d}g_{XR}}{\mathrm{d}\mu} = & \frac{1}{(4\pi)^2}\left(\frac{33}{8} g_{XX}^{2} g_{XR} + \frac{53}{12} g_{RR}^{2} g_{XR} - \frac{\sqrt{6}}{12} g_{RR} g_{XR}^{2} + \frac{33}{8} g_{XR}^{3}\right.\\
& \left. - \frac{\sqrt{6}}{24} g_{RR} g_{XX}^{2} + \frac{53}{12} g_{XX} g_{RR} g_{RX} - \frac{\sqrt{6}}{24} g_{XX} g_{RX} g_{XR}\right)\\
&+\frac{1}{(4\pi)^4}\left( \frac{9}{2}  g_{2L}^{2}g_{XR}^{3} + 4 g_{3C}^{2} g_{XR}^{3} + \frac{17}{4} g_{RR}^{4} g_{XR} + \frac{15}{8} g_{RX}^{2} g_{XR}^{3} + \frac{65}{16} g_{XR}^{5} \right. \\
&+3 g_{2L}^{2} g_{RR}^{2} g_{XR} + 4 g_{3C}^{2} g_{XX}^{2} g_{XR} + \frac{65}{16} g_{XX}^{4} g_{XR} + \frac{65}{8} g_{XX}^{2} g_{XR}^{3} + \frac{45}{4} g_{RR}^{2} g_{XR}^{3}\\
&+12 g_{3C}^{2} g_{RR}^{2} g_{XR} - \frac{3  \sqrt{6}}{16}g_{RR} g_{XX}^{4} - \frac{\sqrt{6}}{2} g_{RR}^{3} g_{XR}^{2} - \frac{3  \sqrt{6}}{4}g_{RR} g_{XR}^{4}\\
&+\frac{9}{2} g_{2L}^{2} g_{XX}^{2} g_{XR} - \frac{\sqrt{6}}{8} g_{XX}^{2} g_{RR}^{3} + \frac{15}{2} g_{XX}^{2} g_{RR}^{2} g_{XR} + \frac{17}{4} g_{XX} g_{RR} g_{RX}^{3}\\
&+\frac{45}{8} g_{RR} g_{XX}^{3} g_{RX} + \frac{45}{8} g_{XX}^{2} g_{RX}^{2} g_{XR} + \frac{17}{4} g_{XX} g_{RR}^{3} g_{RX} + \frac{17}{4} g_{RR}^{2} g_{RX}^{2} g_{XR}\\
&+3 g_{2L}^{2} g_{XX} g_{RR} g_{RX} + 12 g_{3C}^{2} g_{XX} g_{RR} g_{RX} - \frac{15\sqrt{6} }{16} g_{RR} g_{XX}^{2} g_{XR}^{2} \\
& - \frac{9\sqrt{6}}{16} g_{XX} g_{RX} g_{XR}^{3} - \frac{9 \sqrt{6}}{16} g_{XX}^{3} g_{RX} g_{XR} - \frac{3  \sqrt{6}}{8}g_{RR} g_{XX}^{2} g_{RX}^{2}\\
& - \frac{\sqrt{6}}{4} g_{RR} g_{RX}^{2} g_{XR}^{2} - \frac{5 \sqrt{6} }{8} g_{XX}g_{RR}^{2} g_{RX} g_{XR} + \frac{105}{8} g_{XX} g_{RR} g_{RX} g_{XR}^{2} \\
& \left. - \frac{\sqrt{6}}{8} g_{XX} g_{RX}^{3} g_{XR}\right) \ ,
\end{align*}
\begin{align*}
\mu \frac{\mathrm{d}g_{XX}}{\mathrm{d}\mu} = & \frac{1}{(4\pi)^2}\left(\frac{33 }{8}g_{XX}^{3} - \frac{\sqrt{6}}{12} g_{RX} g_{XX}^{2} + \frac{53}{12} g_{XX} g_{RX}^{2} + \frac{33}{8} g_{XX} g_{XR}^{2}\right. \\
& \left. - \frac{\sqrt{6}}{24} g_{XX} g_{RR} g_{XR} + \frac{53}{12} g_{RR} g_{RX} g_{XR} - \frac{\sqrt{6}}{24} g_{RX} g_{XR}^{2}\right)\\ & +\frac{1}{(4\pi)^4}\left(\frac{9}{2} g_{2L}^{2} g_{XX}^{3} + 4 g_{3C}^{2} g_{XX}^{3} + \frac{65}{16} g_{XX}^{5} + \frac{15}{8} g_{XX}^{3} g_{RR}^{2} \right. \\
& + \frac{45}{4} g_{XX}^{3} g_{RX}^{2} + \frac{17}{4} g_{XX} g_{RX}^{4} + 3 g_{2L}^{2} g_{XX} g_{RX}^{2} \\
& + 12 g_{3C}^{2} g_{XX} g_{RX}^{2} + 4 g_{3C}^{2} g_{XX} g_{XR}^{2} + \frac{65}{8} g_{XX}^{3} g_{XR}^{2} + \frac{65}{16} g_{XX} g_{XR}^{4}\\
&- \frac{3 \sqrt{6}}{4} g_{XX}^{4} g_{RX} - \frac{\sqrt{6}}{2} g_{XX}^{2} g_{RX}^{3} - \frac{\sqrt{6}}{8} g_{RX}^{3} g_{XR}^{2} - \frac{3 \sqrt{6}}{16} g_{RX} g_{XR}^{4}\\
&+\frac{9}{2} g_{XX} g_{2L}^{2} g_{XR}^{2} + \frac{17}{4} g_{XX} g_{RR}^{2} g_{RX}^{2} + \frac{15}{2} g_{XX} g_{RX}^{2} g_{XR}^{2} + \frac{17}{4} g_{RR} g_{RX}^{3} g_{XR}\\
&+3 g_{2L}^{2} g_{RR} g_{RX} g_{XR} + \frac{45}{8} g_{XX} g_{RR}^{2} g_{XR}^{2} + \frac{17}{4} g_{RR}^{3} g_{RX} g_{XR} + \frac{45}{8} g_{RR} g_{RX} g_{XR}^{3}\\
&+12 g_{3C}^{2} g_{RR} g_{RX} g_{XR} - \frac{9 \sqrt{6}}{16} g_{RR} g_{XX}^{3} g_{XR} - \frac{15 \sqrt{6}}{16} g_{XX}^{2} g_{RX} g_{XR}^{2}\\
& - \frac{9 \sqrt{6}}{16} g_{XX} g_{RR} g_{XR}^{3}- \frac{\sqrt{6}}{4}g_{RX} g_{XX}^{2} g_{RR}^{2} - \frac{\sqrt{6}}{8} g_{XX} g_{RR}^{3} g_{XR} \\
&\left. - \frac{3 \sqrt{6}}{8} g_{RR}^{2} g_{RX} g_{XR}^{2} +\frac{105}{8} g_{RR} g_{XX}^{2} g_{RX} g_{XR} - \frac{5 \sqrt{6}}{8} g_{XX} g_{RR} g_{RX}^{2} g_{XR}\right) \ .
\end{align*}
\begin{align*} 
{\rm From} \ M_I \ {\rm to} \ M_{X} \ : & \;\;\;b_{2L} =-3, \; b_{2R} =\frac{26}{3}, \; b_{4C} =-\frac{17}{3},  \;\;\; b_{ij}= 
\begin{pmatrix}
8 & 3 & \frac{45}{2} \\ 
3 & \frac{1004}{3} & \frac{1245}{2} \\ 
\frac{9}{2} & \frac{249}{2} & \frac{1315}{6}
\end{pmatrix}.
\end{align*}

\subsection*{A.2\quad The RGEs and $\beta$-coefficients for the Breaking Chain in 
Sec.~\ref{SO10_B}}

From $M_{II}$ to $M_I$:\\
\begin{align*}
\mu \frac{\mathrm{d}g_{2L}}{\mathrm{d}\mu} = &\frac{1}{(4\pi)^2}\left(- 3 g_{2L}^{3}\right) + \frac{1}{(4\pi)^4}\left( 8 g_{2L}^{5} + 12 g_{2L}^{3} g_{3C}^{2} + \frac{3 g_{2L}^{3}}{2} g_{XX}^{2} + g_{2L}^{3} g_{R}^{2} \right. \\
& \left. + g_{2L}^{3} g_{RX}^{2} + \frac{3 g_{2L}^{3}}{2} g_{XR}^{2}\right) \ ,
\end{align*}

\begin{align*}
\mu \frac{\mathrm{d}g_{3C}}{\mathrm{d}\mu} = &\frac{1}{(4\pi)^2}\left(- 7 g_{3C}^{3}\right) + \frac{1}{(4\pi)^4}\left( \frac{9 g_{2L}^{2}}{2} g_{3C}^{3} - 26 g_{3C}^{5} + \frac{g_{3C}^{3}}{2} g_{XX}^{2} + \frac{3 g_{3C}^{3}}{2} g_{R}^{2} \right. \\ 
& \left. + \frac{3 g_{3C}^{3}}{2} g_{RX}^{2} + \frac{g_{3C}^{3}}{2} g_{XR}^{2}\right) \ ,
\end{align*}

\begin{align*}
\mu \frac{\mathrm{d}g_{RR}}{\mathrm{d}\mu} = &\frac{1}{(4\pi)^2}\left(\frac{14 g_{RR}^{3}}{3} - \frac{\sqrt{6}}{3} g_{XR} g_{RR}^{2} + \frac{14}{3} g_{RR} g_{RX}^{2} + \frac{9}{2} g_{RR} g_{XR}^{2}\right. \\
& \left. - \frac{g_{XX}}{6} g_{RR} \sqrt{6} g_{RX} + \frac{9}{2} g_{XX} g_{RX} g_{XR} - \frac{g_{XR}}{6} \sqrt{6} g_{RX}^{2}\right) \\
&+ \frac{1}{(4\pi)^4}\left( 3 g_{2L}^{2} g_{RR}^{3} + 12 g_{3C}^{2} g_{RR}^{3} + \frac{15}{2} g_{XX}^{2} g_{RR}^{3} + 8 g_{RR}^{5} + 16 g_{RR}^{3} g_{RX}^{2} \right.\\ 
&+ 45 g_{RR}^{3} g_{XR}^{2} + 8 g_{RR} g_{RX}^{4} - 8 \sqrt{6} g_{RR}^{4} g_{XR} - 12 \sqrt{6} g_{RR}^{2} g_{XR}^{3}\\
& + \frac{25}{2} g_{RR} g_{XR}^{4} - 2 \sqrt{6} g_{RX}^{4} g_{XR} - 3 \sqrt{6} g_{RX}^{2} g_{XR}^{3}+3 g_{2L}^{2} g_{RR} g_{RX}^{2} \\
&+ \frac{9}{2} g_{RR} g_{2L}^{2} g_{XR}^{2} + 12 g_{3C}^{2} g_{RR} g_{RX}^{2} + 4 g_{3C}^{2} g_{RR} g_{XR}^{2} + 30 g_{RR} g_{RX}^{2} g_{XR}^{2}\\
&+\frac{25 g_{XX}^{3}}{2} g_{RX} g_{XR} + \frac{25}{2} g_{RR} g_{XX}^{2} g_{XR}^{2} + \frac{45}{2} g_{XX} g_{RX}^{3} g_{XR} \\
&+ \frac{25}{2} g_{XX} g_{RX} g_{XR}^{3}+\frac{45}{2} g_{RR} g_{XX}^{2} g_{RX}^{2} - 9 \sqrt{6} g_{XX}^{2} g_{RX}^{2} g_{XR}\\
& - 6 \sqrt{6} g_{XX} g_{RR} g_{RX}^{3} - 10 \sqrt{6} g_{RR}^{2} g_{RX}^{2} g_{XR} + 4 g_{3C}^{2} g_{XX} g_{RX} g_{XR} \\
& - 3 \sqrt{6} g_{XX}^{3} g_{RR} g_{RX} - 6 \sqrt{6} g_{XX}^{2} g_{RR}^{2} g_{XR} - 6 \sqrt{6} g_{XX} g_{RR}^{3} g_{RX}\\
& \left. + \frac{9}{2} g_{XX} g_{2L}^{2} g_{RX} g_{XR} + \frac{105}{2} g_{XX} g_{RR}^{2} g_{RX} g_{XR} - 15 \sqrt{6} g_{XX} g_{RR} g_{RX} g_{XR}^{2}\right) \ ,
\end{align*}

\begin{align*}
\mu \frac{\mathrm{d}g_{RX}}{\mathrm{d}\mu} = &\frac{1}{(4\pi)^2}\left(\frac{9}{2} g_{XX}^{2} g_{RX} - \frac{\sqrt{6}}{3} g_{XX} g_{RX}^{2} + \frac{14}{3} g_{RR}^{2} g_{RX} + \frac{14}{3} g_{RX}^{3}\right. \\
& \left. - \frac{\sqrt{6}}{6} g_{XX} g_{RR}^{2} + \frac{9}{2} g_{XX} g_{RR} g_{XR} - \frac{\sqrt{6}}{6} g_{RR} g_{RX} g_{XR}\right)\\
& + \frac{1}{(4\pi)^4}\left(3 g_{2L}^{2} g_{RX}^{3} + 12 g_{3C}^{2} g_{RX}^{3} + 45 g_{XX}^{2} g_{RX}^{3} + 8 g_{RR}^{4} g_{RX} + 16 g_{RR}^{2} g_{RX}^{3} \right. \\
&+ 8 g_{RX}^{5} + \frac{15}{2} g_{RX}^{3} g_{XR}^{2}+\frac{25 g_{XX}^{4}}{2} g_{RX} - 3 \sqrt{6} g_{XX}^{3} g_{RR}^{2} - 12 \sqrt{6} g_{XX}^{3} g_{RX}^{2} \\
&- 2 \sqrt{6} g_{XX} g_{RR}^{4} - 8 \sqrt{6} g_{XX} g_{RX}^{4} + \frac{9}{2} g_{2L}^{2} g_{XX}^{2} g_{RX} + 3 g_{2L}^{2} g_{RR}^{2} g_{RX}  \\
& + 4 g_{3C}^{2} g_{XX}^{2} g_{RX} + 12 g_{3C}^{2} g_{RR}^{2} g_{RX} + 30 g_{XX}^{2} g_{RR}^{2} g_{RX}+\frac{25}{2} g_{RR} g_{XX}^{3} g_{XR}\\
& + \frac{25}{2} g_{XX}^{2} g_{RX} g_{XR}^{2} + \frac{45}{2} g_{XX} g_{RR}^{3} g_{XR} + \frac{25}{2} g_{XX} g_{RR} g_{XR}^{3}\\
&- 10 \sqrt{6} g_{XX} g_{RR}^{2} g_{RX}^{2} - 9 \sqrt{6} g_{XX} g_{RR}^{2} g_{XR}^{2} - 6 \sqrt{6} g_{XX} g_{RX}^{2} g_{XR}^{2} \\
& + \frac{45}{2} g_{RR}^{2} g_{RX} g_{XR}^{2} + 4 g_{3C}^{2} g_{XX} g_{RR} g_{XR} - 6 \sqrt{6} g_{RR}^{3} g_{RX} g_{XR}\\
& - 6 \sqrt{6} g_{RR} g_{RX}^{3} g_{XR} - 3 \sqrt{6} g_{RR} g_{RX} g_{XR}^{3} + \frac{9}{2} g_{XX} g_{2L}^{2} g_{RR} g_{XR}\\
& \left. - 15 \sqrt{6} g_{XX}^{2} g_{RR} g_{RX} g_{XR} + \frac{105}{2} g_{XX} g_{RR} g_{RX}^{2} g_{XR}\right) \ ,
\end{align*}
\begin{align*}
\mu \frac{\mathrm{d}g_{XR}}{\mathrm{d}\mu} = &\frac{1}{(4\pi)^2}\left(\frac{9}{2} g_{XX}^{2} g_{XR} + \frac{14}{3} g_{RR}^{2} g_{XR} - \frac{\sqrt{6}}{3} g_{RR} g_{XR}^{2} + \frac{9}{2} g_{XR}^{3} \right. \\
& \left. - \frac{\sqrt{6}}{6} g_{RR} g_{XX}^{2} + \frac{14}{3} g_{XX} g_{RR} g_{RX} - \frac{\sqrt{6}}{6} g_{XX} g_{RX} g_{XR}\right)\\
& + \frac{1}{(4\pi)^4}\left(\frac{9 g_{2L}^{2}}{2} g_{XR}^{3} + 4 g_{3C}^{2} g_{XR}^{3} + 25 g_{XX}^{2} g_{XR}^{3} + 8 g_{RR}^{4} g_{XR} \right. \\
& + 45 g_{RR}^{2} g_{XR}^{3} + \frac{25}{2} g_{XR}^{5} - 3 \sqrt{6} g_{XX}^{4} g_{RR} + \frac{25 g_{XX}^{4}}{2} g_{XR} - 8 \sqrt{6} g_{RR}^{3} g_{XR}^{2} \\
& - 12 \sqrt{6} g_{RR} g_{XR}^{4} + \frac{15}{2} g_{RX}^{2} g_{XR}^{3} + 3 g_{2L}^{2} g_{RR}^{2} g_{XR} + 4 g_{3C}^{2} g_{XX}^{2} g_{XR}\\
& - 2 \sqrt{6} g_{XX}^{2} g_{RR}^{3} + 8 g_{XX} g_{RR}^{3} g_{RX} + 8 g_{XX} g_{RR} g_{RX}^{3} + 8 g_{RR}^{2} g_{RX}^{2} g_{XR}\\
&+\frac{9 g_{2L}^{2}}{2} g_{XX}^{2} g_{XR} + 12 g_{3C}^{2} g_{RR}^{2} g_{XR} + \frac{45}{2} g_{RR} g_{XX}^{3} g_{RX} + 30 g_{XX}^{2} g_{RR}^{2} g_{XR}\\
&- 9 \sqrt{6} g_{XX}^{3} g_{RX} g_{XR} - 15 \sqrt{6} g_{XX}^{2} g_{RR} g_{XR}^{2} + \frac{45}{2} g_{XX}^{2} g_{RX}^{2} g_{XR} \\
&- 9 \sqrt{6} g_{XX} g_{RX} g_{XR}^{3}+3 g_{2L}^{2} g_{XX} g_{RR} g_{RX} - 6 \sqrt{6} g_{XX}^{2} g_{RR} g_{RX}^{2} \\
& - 2 \sqrt{6} g_{XX} g_{RX}^{3} g_{XR} - 4 \sqrt{6} g_{RR} g_{RX}^{2} g_{XR}^{2} +12 g_{3C}^{2} g_{XX} g_{RR} g_{RX}\\
&\left. - 10 \sqrt{6} g_{XX} g_{RR}^{2} g_{RX} g_{XR} + \frac{105}{2} g_{XX} g_{RR} g_{RX} g_{XR}^{2}\right) \ ,
\end{align*}
\begin{align*}
\mu \frac{\mathrm{d}g_{XX}}{\mathrm{d}\mu} = &\frac{1}{(4\pi)^2}\left(\frac{9 g_{XX}^{3}}{2} - \frac{g_{RX}}{3} \sqrt{6} g_{XX}^{2} + \frac{14}{3} g_{XX} g_{RX}^{2} + \frac{9}{2} g_{XX} g_{XR}^{2} \right. \\
& \left. - \frac{ \sqrt{6} }{6} g_{RR}g_{XX}g_{XR} + \frac{14}{3} g_{RR} g_{RX} g_{XR} - \frac{\sqrt{6}}{6} g_{RX} g_{XR}^{2}\right)\\
& + \frac{1}{(4\pi)^4}\left(\frac{9 g_{2L}^{2}}{2} g_{XX}^{3} + 4 g_{3C}^{2} g_{XX}^{3} + \frac{25 g_{XX}^{5}}{2} + 45 g_{XX}^{3} g_{RX}^{2} \right. \\
& + 25 g_{XX}^{3} g_{XR}^{2} + 8 g_{XX} g_{RX}^{4} - 12 \sqrt{6} g_{XX}^{4} g_{RX} + \frac{15}{2} g_{XX}^{3} g_{RR}^{2}\\
& - 8 \sqrt{6} g_{XX}^{2} g_{RX}^{3} + \frac{25}{2} g_{XX} g_{XR}^{4} - 3 \sqrt{6} g_{RX} g_{XR}^{4} + 3 g_{2L}^{2} g_{XX} g_{RX}^{2}\\
& + 4 g_{3C}^{2} g_{XX} g_{XR}^{2} + 8 g_{XX} g_{RR}^{2} g_{RX}^{2} + 8 g_{RR} g_{RX}^{3} g_{XR} - 2 \sqrt{6} g_{RX}^{3} g_{XR}^{2}\\
&+\frac{9}{2} g_{XX} g_{2L}^{2} g_{XR}^{2} + 12 g_{3C}^{2} g_{XX} g_{RX}^{2} + \frac{45}{2} g_{XX} g_{RR}^{2} g_{XR}^{2} + 30 g_{XX} g_{RX}^{2} g_{XR}^{2} \\
& + 8 g_{RR}^{3} g_{RX} g_{XR} - 9 \sqrt{6} g_{XX}^{3} g_{RR} g_{XR} - 15 \sqrt{6} g_{XX}^{2} g_{RX} g_{XR}^{2} \\
& - 9 \sqrt{6} g_{XX} g_{RR} g_{XR}^{3} + \frac{45}{2} g_{RR} g_{RX} g_{XR}^{3} + 3 g_{2L}^{2} g_{RR} g_{RX} g_{XR} \\
& - 4 \sqrt{6} g_{XX}^{2} g_{RR}^{2} g_{RX} - 2 \sqrt{6} g_{XX} g_{RR}^{3} g_{XR} - 6 \sqrt{6} g_{RR}^{2} g_{RX} g_{XR}^{2}\\
&\left. +12 g_{3C}^{2} g_{RR} g_{RX} g_{XR} + \frac{105}{2} g_{RR} g_{XX}^{2} g_{RX} g_{XR} - 10 \sqrt{6} g_{XX} g_{RR} g_{RX}^{2} g_{XR}\right) \ .
\end{align*}

\begin{align*} 
{\rm From} \ M_I \ {\rm to} \ M_{X} \ : & \;\;\;b_{2L} =-3, \; b_{2R} =\frac{8}{3}, \; b_{4C} =-\frac{25}{3},  \;\;\; b_{ij}= 
\begin{pmatrix}
8 & 3 & \frac{45}{2} \\ 
3 & \frac{470}{3} & \frac{555}{2} \\ 
\frac{9}{2} & \frac{111}{2} & \frac{130}{3}
\end{pmatrix}.
\end{align*}

\section*{}
\providecommand{\href}[2]{#2}
\addcontentsline*{toc}{section}{}
\bibliographystyle{JHEP}
\bibliography{GUT_TD}
\end{document}